\documentclass[11pt]{article}

\usepackage{amsfonts,amsthm}
\usepackage{amsmath}  
\usepackage{latexsym}
\usepackage{amssymb}
\usepackage{mathrsfs}
\usepackage{tikz}

\newtheorem{lem}{Lemma}[section]
\newtheorem{prop}{Proposition}[section]

\newtheorem{rem}{Remark}

\def\be{\begin{equation}}
\def\ee{\end{equation}}
\def\bea{\begin{eqnarray}}
\def\eea{\end{eqnarray}}

\newcommand{\ie}{{\it i.e.}\ }
\newcommand{\lda}{ w}
\newcommand{\Ad}{\mathbb{A}}
\newcommand{\HH}{\mathbb{H}}
\newcommand{\RR}{\mathbb{R}}
\newcommand{\CC}{\mathbb{C}}
\newcommand{\ZZ}{{\mathbb{Z}}}

\newcommand{\II}{\mathbb{I}}

\newcommand{\MM}{\mathbb{M}}

\newcommand{\cL}{\mathscr{L}}
\newcommand{\cM}{\mathscr{M}}
\newcommand{\cA}{\mathscr{A}}

\newcommand{\ct}{\mathfrak{t}}
\newcommand{\cC}{\mathcal C}

\newcommand{\cI}{\mathcal I}
\newcommand{\1}{\mathbb{I}}
\newcommand{\prf}{\underline{Proof:}\ }
\newcommand{\finprf}{\null \hfill {\rule{5pt}{5pt}}\\ \null}

\numberwithin{equation}{section}
\begin{document}

\begin{center}
	\textbf{\Large New integrable boundary conditions for the Ablowitz--Ladik model: from Hamiltonian formalism to nonlinear mirror image method}\\[3ex]
	\large{Vincent Caudrelier$^a$, Nicolas Cramp\'e$^{b,c,d}$}\\[3ex]

	$^a$ School of Mathematics, University of Leeds, LS2 9JT, UK  \\[3ex]
	
	$^b$ Laboratoire Charles Coulomb (L2C), Univ Montpellier, CNRS, Montpellier, France.\\[3ex]
	
	$^c$ Centre de Recherches Math\'ematiques, Universit\'e de Montr\'eal, Montr\'eal (QC), Canada.\\[3ex]
	
	$^d$ Institut Denis-Poisson - Universit\'e de Tours - Universit\'e d’Orl\'eans, Tours, France.\\[3ex]
	
\end{center}

\vspace{0.5cm}

\centerline{\bf Abstract}  
Using Sklyanin's classical theory of integrable boundary conditions, we use the Hamiltonian approach to derive new integrable boundary conditions for the Ablowitz--Ladik model on the finite and half infinite lattice. In the case of half infinite lattice, the special and new emphasis of this paper is to connect directly the Hamiltonian approach, based on the classical $r$-matrix, with the zero curvature representation and B\"acklund transformation approach that allows one to implement a nonlinear mirror image method and construct explicit solutions. It is shown that for our boundary conditions, which generalise (discrete) Robin boundary conditions, a nontrivial extension of the known mirror image method to what we call {\it time-dependent boundary conditions} is needed. A careful discussion of this extension is given and is facilitated by introducing the notion of intrinsic and extrinsic picture for describing boundary conditions. This gives the specific link between Sklyanin's reflection matrices and B\"acklund transformations combined with folding, {\it in the case of non-diagonal reflection matrices}. All our results reproduce the known Robin boundary conditions setup as a special case: the diagonal case. Explicit formulas for constructing multisoliton solutions on the half-lattice with our time-dependent boundary conditions are given and some examples are plotted. 

\section{Introduction}

The problem of formulating integrable partial differential equations (PDEs) on finite or half-infinite intervals appeared very soon after the discovery of the Inverse Scattering Method (ISM) \cite{GGKM,ZS}, in \cite{AS}. However, this did not address, in the context of boundary conditions, one of the most important aspects of integrable systems: the fact that integrable PDEs (in $1+1$ dimensions) are (infinite-dimensional) Hamiltonian systems. 
Sklyanin's seminal work \cite{Sk} paved the way to a framework that tackles boundary conditions in integrable classical systems both from the point of view of ISM and of Hamiltonian theory\footnote{We do not mention quantum integrable boundary conditions here. The literature is so vast that even an attempt at surveying the main results would take us to far astray.}. The PDE point of view seems to have prevailed however until recently and developed into the so-called nonlinear mirror image method, following the initial impetus of \cite{H,BT1,BT2,T} and revived more recently e.g. in \cite{BH1,CZ1,CZ2}. Note also a new related angle on the question, called boundary dressing, which appeared in \cite{Z,ZZ}. The key tools of that method are B\"acklund transformation and a special folding symmetry. 

On the one hand, the Hamiltonian approach to integrable boundary conditions rests upon the classical reflection equation which describes the allowed boundary conditions via its solutions: the reflection matrices. The latter are conditioned by the classical $r$-matrix \cite{Skly,STS1} of the model under consideration. On the other hand, the nonlinear mirror image approach relies on the use of a B\"acklund matrix which satisfies what we could call a boundary zero curvature equation. Both aspects are contained in \cite{Sk}. However, a thorough analysis of the class of solutions allowed by one or the other method shows that the reflection equation admit more general solutions (non-diagonal reflection matrices) than the boundary zero curvature equation. This begged the question: why such a discrepancy and how can we resolve it? This question was the motivation behind \cite{ACC} where a detailed answer was provided. The non-diagonal reflection matrices can be cast into a boundary zero curvature form provided a time dependent term is included. This time-dependent term is in fact the time derivative of the reflection matrix itself. In other words, the way out of this discrepancy is to consider a time-dependent version of reflection matrices. This was of course well-known in the Hamiltonian setup where they appear as solution of the so-called dynamical reflection equation \cite{Skquant}. As it turns out, 
the time-dependent boundary zero curvature equation has also appeared before in the literature but its direct connection with Hamiltonian setup was not understood. Rather, some a posteriori checks were performed in some instances (see e.g. \cite{Cor,BCDR} and references therein for this point of view). 

The main aim of the present paper is to follow the general results obtained in \cite{ACC} and illustrate them in full detail on the example of the Ablowitz--Ladik (AL) model \cite{AL}. In doing so, we actually find new integrable boundary conditions for the AL model. In fact, we work with a more general Ablowitz--Ladik-type model which depends on three arbitrary parameters, see e.g. \cite{XF}. For special values of the parameters, one can recover the usual AL model or a discrete modified Korteweg-de Vries model. Our approach is as follows. Starting from the Hamiltonian approach, we obtain the general non-diagonal solution of Sklyanin's classical reflection equation and show how to {\it derive}  the (boundary) zero curvature equations. Our results contain the case of (discrete) Robin boundary conditions treated previously in \cite{BH} as a particular case. Using the method introduced in \cite{AD1}, we compute the Lax pair for the model on the finite lattice. We then make the explicit connection with the nonlinear mirror image method for our new integrable boundary conditions. To facilitate the transition, we introduce the notion of {\it intrinsic and extrinsic picture} for describing integrable boundary conditions. They are related by a local gauge transformation which realises the transition from the time-independent form of the boundary zero curvature equation (intrinsic picture) to its time-dependent form (extrinsic picture). Once in the extrinsic picture, we are able to use the ideas of the nonlinear mirror image method and to implement them in our case. Explicit formulas for multisolitons solutions in the presence of our new integrable boundary conditions are provided via symmetries on the scattering data coming from the folding procedure. 

In Section \ref{AL}, we review the classical $r$-matrix approach for the Ablowitz--Ladik (AL) model with periodic boundary conditions in order to introduce the required notations and tools. Note that we obtain different results from the standard ones as we use a more convenient normalisation for the Lax matrix of AL. We recall the notion of monodromy and transfer matrices and how one can {\it derive} the Lax pair and the zero curvature representation of the equations of motions from the Hamiltonian formalism. In Section \ref{AL_openBC}, the necessary modifications to the $r$-matrix approach, as proposed in \cite{Sk}, are implemented for AL on the finite chain with integrable boundary conditions in our normalisation. We derive a general reflection matrix, obtain the corresponding equations of motion on the open interval, first as Hamiltonian equations of motion and then as a zero curvature equation for an appropriately modified Lax pair, to take the boundary conditions into account. 
In section \ref{ALboundary}, we introduce the notion of intrinsic and extrinsic picture for boundary conditions. It is first illustrate for the case of (discrete) Robin boundary conditions and then implemented for our more general case. The time-dependent form of our new boundary conditions is then obtained. In Section \ref{LP}, the Lax pair and (boundary) zero curvature representation of the model with boundary conditions are derived from the Hamiltonian approach. The extrinsic picture is described in this setup, which allows us to make the connection with the nonlinear mirror image method. In Section \ref{mirror}, the results of the previous section are used to construct explicit solutions, in particular multisoliton solutions (see Proposition \ref{form_solitons}), for the problem on the half-lattice with our integrable boundary conditions. We restrict our attention to the discrete NLS reduction in the focusing regime and implement the nonlinear mirror image method. Some conclusions are gathered in the last section. 

\section{ Ablowitz--Ladik model with periodic boundary conditions}\label{AL}

\subsection{Transfer matrix formalism and integrable Hamiltonian}

The classical $r$-matrix approach provides a neat and powerful formalism to present the Hamiltonian formulation of integrable systems. It also allows one to derive the time Lax matrix and the zero curvature representation as Hamilton's equations.
In this section, we follow the well-known method (see e.g. \cite{FT}) and adapt it to the Ablowitz--Ladik model. We stress that the explicit results are actually new since we use a different normalisation of the Lax pair, which turns out to change the $r$-matrix underlying the model. In particular, the Liouville integrability of the general AL system \eqref{eq:qp}-\eqref{eq:rp} below is established for the first time here to the best of our knowledge. The normalisation we use was studied in \cite{XF} where it was advocated to be better than the traditional one as it produces a Lax pair with appealing algebraic properties. We will see throughout this paper that indeed, this normalisation is superior for various reasons. 

The starting point of the classical r-matrix approach to Hamiltonian integrable lattice models is the so-called Lax matrix 
$\ell(j,z)$, where $j$ is an integer associated to a site in the chain and $z$ is the spectral parameter, which is assumed to obey the following quadratic ultralocal Poisson algebra
\begin{equation}
\label{rll}
 \{\ell_a(j, w)\ ,\ \ell_b(k, z)\}=\delta_{jk}\left[\ r_{ab}( w/z)\ , \ \ell_a(j, w) \ell_b(k, z) \ \right]\,,
\end{equation}
which we consider in $\CC^2\otimes\CC^2$ here. This Poisson algebra encodes the Poisson brackets used on the phase space of the model. 
The standard auxiliary space notation has been used here
\be
\ell_a=\ell\otimes \1\,,~~\ell_b=\1\otimes\ell\,,~~\{\ell_a(j, w)\ ,\, \ell_b(k, z)\}=\sum_{m,n,p,q=1}^2\{\ell^{mn}(j, w)\ ,\ \ell^{pq}(k, z)\}\,E_{mn}\otimes E_{pq}\,,
\ee 
where $\1$ is the $2\times 2$ identity matrix, $E_{mn}$ is the canonical basis of $2\times 2$ matrices and $\ell^{mn}(j, w)$ are the entries of the matrix $\ell(j,z)$.
The classical $r$-matrix $r_{ab}$ is a $4\times 4$ matrix. For the Ablowiz-Ladik model we take 
\begin{eqnarray}
\label{normalisation_ell}
\ell(j, z)=\frac{1}{\sqrt{1-q_jr_j}}\begin{pmatrix}
z& q_j\\
r_j& 1/ z
\end{pmatrix}\;.
\end{eqnarray}
Notice the extra factor in front of the matrix which ensures that $\det \ell(j,z)=1$. With this choice, one can check that having the 
$r$-matrix in \eqref{rll} of the form
\begin{equation}\label{eq:r}
 r( z)=\frac{i}{2(1- z^2)}\begin{pmatrix}
        z^2+1 & 0 & 0 & 0\\
       0&0 &2 z&0\\
       0&2 z&0 &0\\
       0 & 0 &0 &  z^2+1
      \end{pmatrix}
\end{equation}
is equivalent to the following Poisson brackets on the fields
\begin{equation}
\label{PB_fields}
 \{q_j\ , \ q_k \}=\{r_j\ , \ r_k \}=0\quad\text{and}\qquad \{q_j\ , \ r_k \}=i\delta_{jk}(1-q_kr_k)\;.
\end{equation}
These are the standard AL brackets.
The r-matrix \eqref{eq:r} is skew-symmetric 
\be
r_{ab}(\lda)=-r_{ba}(1/\lda)
\ee
and satisfies the classical Yang-Baxter equation
\begin{eqnarray}
 [\ r_{ac}(\lda/\nu)\ ,\ r_{bc}( z/\nu)\ ]+ [\ r_{ab}(\lda/ z)\ ,\ r_{ac}(\lda/\nu)\ ]+ [\ r_{ab}(\lda/ z)\ ,\ r_{bc}( z/\nu)\ ]=0\,.
\end{eqnarray}
These two properties ensure that the Poisson bracket defined by \eqref{rll} is antisymmetric and satisfies the Jacobi property.
In addition to these properties, the r-matrix \eqref{eq:r} is symmetric in the auxiliary spaces
\be
\label{sym}
r_{ab}(\lda)=r_{ba}(\lda)\,.
\ee

\begin{rem}
There is a one-parameter family of normalisations for $\ell(j,z)$:
 \begin{eqnarray}
 \ell^{(s)}(j, z)=(1-q_jr_j)^s\begin{pmatrix}
             z& q_j\\
            r_j& 1/z
           \end{pmatrix}\;,
\end{eqnarray}
for all $s\in\RR$. It is still possible to find an $r$-matrix for each $s$
\begin{equation}\label{eq:r2}
 r^{(s)}(z)=\frac{i}{2(1- z^2)}\begin{pmatrix}
        z^2+1 & 0 & 0 & 0\\
       0&(1+2s)( z^2-1) &2 z&0\\
       0&2 z&(1+2s)(1- z^2)&0\\
       0 & 0 &0 &  z^2+1
      \end{pmatrix}\;.
\end{equation}
The case $s=0$ corresponds to the normalisation which is most often used for the AL model. 
In the present paper, we use $s=-1/2$ which is the particular value for which $r^{(s)}$ simplifies nicely and enjoys the additional property \eqref{sym} of being symmetric. This 
is another argument in favour of this particular normalisation, in addition to those advocated in \cite{XF}.
\end{rem}

The Hamiltonian of the model on $N+1$ sites (and all conserved quantities) can be obtained by defining the so-called single-row monodromy matrix
\begin{equation}
\label{single_mono}
 L_a(z)=\ell_a(N,z)\ell_a(N-1,z)\dots \ell_a(1,z)\ell_a(0,z)\,,
\end{equation}
and the associated single-row transfer matrix
\begin{equation}
\label{single_row}
\ct( z)=tr_a L_a(z)
\end{equation}
Then, one shows that the following holds
\begin{eqnarray}
\label{rt}
&& \{L_a( w), L_b( z)\}=\left[ r_{ab}( w/ z)\ , \ L_a( w) \ L_b( z) \right]\,,\\
\label{PB_commute}&& \{\ct(  w) , \ct( z)\}=0\;.
\end{eqnarray}
The second relation allows us to take $\ct( w)$ as the generating function in $ w$ of elements $\cI^{(n)}$ in involution:
\begin{eqnarray}
\label{expansion_t}
 \ct(z) = \sum_{j=0}^{N+1}  \cI^{(-N+2j-1)}  z^{-N+2j-1}
\end{eqnarray}
Let us introduce
\begin{equation}
 \cC=\prod_{j=0}^N \frac{1}{\sqrt{1-q_jr_j}}\;.
\end{equation}
Then, by direct calculation, one shows that
\begin{eqnarray}
&& \cI^{(N+1)}=\cI^{(-N-1)}=\cC\;,\\
&& \cI^{(N-1)}= \cC \sum_{j=0}^{N} r_jq_{j+1}\qquad\text{and} \quad\cI^{(-N+1)}= \cC \sum_{j=0}^{N} q_jr_{j+1}\,
\end{eqnarray}
where we have used the conventions $q_{N+1}\equiv q_0$ and $r_{N+1}\equiv r_0$.

Let us introduce the Ablowitz--Ladik type Hamiltonian as the following combination of elements in involution
\begin{eqnarray}
\label{AL_Ham}
H&=&-2\alpha\cC^{-1}\cI^{(N-1)}-2\beta\cC^{-1}\cI^{(-N+1)}-4 \gamma \ln \cC\nonumber\\
&=&2\sum_{j=0}^{N}\big(- \alpha r_jq_{j+1}-\beta q_jr_{j+1} + \gamma \ln(1-q_jr_j) \big)\;. \qquad
\end{eqnarray}
We associate to the Hamiltonian $H$ a time evolution according to 
\begin{equation}
 \partial_{t}\ \boldsymbol{\cdot}=\{ H\ ,\ \boldsymbol{\cdot} \}\,.
\end{equation}
With this choice, the equations of motion introduced in \cite{XF} appear as Hamilton's equations for the fields $q_j$, $r_j$ contained in $\ell(j,z)$ 
\begin{equation}
\label{eq_Hamilton}
\partial_{t}\ \ell(j,z)=\{ H ,\ \ell(j,z) \}\,,
\end{equation}
\ie explicitly, for $j=0,1,\dots,N$,
\begin{eqnarray}
&&\dot{q}_j=2i\big(\alpha q_{j+1}+\gamma q_j+\beta q_{j-1} -q_jr_j(\alpha q_{j+1}+\beta q_{j-1})\big)\label{eq:qp} \\
&&\dot{r}_j=-2i\big(\beta r_{j+1}+\gamma r_j+\alpha r_{j-1} -q_jr_j(\alpha r_{j-1}+\beta r_{j+1})\big)\label{eq:rp}\,,
\end{eqnarray}
together with periodic boundary conditions $q_{N+1}\equiv q_0$, $r_{N+1}\equiv r_0$, $q_{-1}\equiv q_N$ and $r_{-1}\equiv r_N$.
Since these equations of motion derive from the Hamiltonian $H$ which was constructed from the transfer matrix $\ct(z)$, this proves the Liouville integrability of the model with periodic boundary conditions.

The Ablowitz--Ladik equations are recovered for $\alpha=\beta=\frac{1}{2}$ and  $\gamma=-1$. With the additional reduction $r_j=\nu q_j^*$ ($\nu=\pm 1$), 
equations of motion \eqref{eq:qp} and \eqref{eq:rp}
becomes the ones of the (integrable) discrete nonlinear Schr\"odinger equation
\begin{equation}
 \dot{q}_j=i\big( q_{j+1}-2 q_j+ q_{j-1} -\nu |q_j|^2( q_{j+1}+ q_{j-1})\big)\;.\label{eq:DNLS}
\end{equation}
For $\alpha=-\beta=\frac{i}{2}$ and $\gamma=0$ with the reduction $q_j=\nu r_j$ (and $\nu=\pm 1$ and  $q_j$ real), we recover the equations of motion of the discrete modified Korteweg-de Vries equation given by
\begin{equation}
 \dot{q}_j=q_{j-1}- q_{j+1}  +\nu q_j^2( q_{j+1}- q_{j-1})\;.\label{eq:dkdv}
\end{equation}

\subsection{Lax pair associated to the Ablowitz--Ladik chain}

It is one of the remarkable features of the $r$-matrix approach that instead of guessing a Lax pair for a given system of equations, 
one can {\it derive} the time Lax matrix $A(j, z)$ from the knowledge of the space Lax matrix $\ell(j,z)$ and its associated 
$r$-matrix\footnote{In fact, one can derive all the Lax matrices $A^{(n)}(j, z)$ corresponding to the commuting higher flows 
but that will not be used here}. The zero-curvature form of the equations of motion (or compatibility condition of the Lax pair) 
is also a by product of this approach. We implement this for the AL chain in the present normalisation. In particular, we will derive the Lax matrix $A(j, z)$ given in \cite{XF}.
 
Let us define the partial monodromy for $n\geq m$
\begin{equation}
L_a(n,m, z)=\ell_a(n, z)\ell_a(n-1, z)\dots \ell_a(m, z)\;.
\end{equation}
We use the convention $L(n-1,n, z)=\1$ and obviously one gets $L(N,0, z)=L(z)$. Following \cite{STS1,Fad}, one defines, for $j=0,\dots,N+1$,
\begin{equation}
\label{Mbar}
 \overline{M}_b(j, w, z)=tr_a\big(\ L_a(N,j, w)\ r_{ab}( w- z)\ L_a(j-1,0, w)\ \big)\;.
\end{equation}
To simplify the expansion in terms of the spectral parameter $w$, we introduce a regularized version of $\overline{M}_b(j, w, z)$ as
\begin{equation}
\label{M}
 {M}(j, w, z)=\overline{M}(j, w, z)-\frac{1}{ w- z}\text{res}_{ w= z}\overline{M}(j, w, z)-\frac{1}{ w+ z}\text{res}_{ w=- z}  \overline{M}(j, w, z)\;.
\end{equation}
The matrix $M(j, w, z)$ is the generating function in $w$ of the matrices $M^{(n)}(j, z)$
\begin{eqnarray}
 M(j, w, z) = \sum_{j=0}^{N+1}  M^{(-N+2j-1)}(j, z)  w^{-N+2j-1}\;.
\end{eqnarray}
In particular, one gets 
\begin{eqnarray}
 M^{(-N-1)}(j, z)&=& \frac{i}{2}\begin{pmatrix}
                     0 & 0\\
                     0 & \cC
                    \end{pmatrix}
 \label{eq:M1} \\
 M^{(-N+1)}(j, z)&=& \frac{i\cC}{2} \begin{pmatrix}
                     q_{j-1}r_j & 2q_{j-1}\\
                     2r_j & 2+ z^2\sum_{p=0}^{N} q_p r_{p+1}
                    \end{pmatrix}\\
 M^{(N-1)}(j, z)&=& -\frac{i\cC}{2}\begin{pmatrix}
                                 2 z^2+\sum_{p=0}^{N} r_p q_{p+1} &2 z q_j\\
                             2   z r_{j-1}    &r_{j-1}q_j
                                  \end{pmatrix}
\\
 M^{(N+1)}(j, z)&=&  -\frac{i}{2}\begin{pmatrix}
                     \cC & 0\\
                     0 & 0
                    \end{pmatrix} \label{eq:M4}
\end{eqnarray}
Using relation \eqref{rll}, one proves
\begin{equation}
\label{ZC_alg}
 \{ \ct( w),\ell(j, z) \}=M(j+1, w, z)\ell(j, z) - \ell(j, z) M(j, w, z)\;.
\end{equation}
Comparing with the expansion \eqref{expansion_t} for $\ct$, this shows how the matrices $M^{(n)}(j, z)$ are associated to the charges  $\cI^{(n)}$. Indeed, by expanding relation \eqref{ZC_alg} with respect to $ w$, one gets
\begin{equation}\label{eq:IM}
 \{ \cI^{(n)},\ell(j, z) \}=M^{(n)}(j+1, z)\ell(j, z) - \ell(j, z) M^{(n)}(j, w, z)\;.
\end{equation}
\begin{rem}
We see that the explicit form of the matrices $M^{(-N-1)}(j, z)$ and $M^{(N+1)}(j, z)$ given by \eqref{eq:M1} and \eqref{eq:M4} are different whereas the corresponding charges  
$\cI^{(-N-1)}$ and $\cI^{(N+1)}$ are equal to $\cC$. This apparent contradiction is solved by noting that we can always add to any 
$M^{(n)}(j, z)$ a matrix proportional to the identity matrix and independent of the site $j$.
In the following, we will use $M^{(-N-1)}(j, z)$ or $M^{(N+1)}(j, z)$ for the charges $\cC$.
\end{rem}

In particular, in view of the expression \eqref{AL_Ham} of $H$ in terms of the $\cI^{(n)}$'s, we obtain
\begin{equation}
 \{ H,\ell(j, z) \}=-2\alpha\{\cC^{-1}\cI^{(N-2)},\ell(j, z) \}-2\beta\{\cC^{-1}\cI^{(-N+2)},\ell(j, z) \}-4 \gamma \{\ln \cC,\ell(j, z) \}\;.
\end{equation}
Using the Leibniz rule and relation \eqref{eq:IM}, this becomes
\begin{equation}\label{ZC_alg2}
 \{ H,\ell(j, z) \}=A(j+1, z)\ell(j, z) - \ell(j, z) A(j, w, z)\;,
\end{equation}
where we have defined
\begin{eqnarray}
 A(j, z)&=&\frac{2}{\cC}\Big( -\alpha M^{(N-1)}(j, z)+ (\alpha\cI^{(N-1)}\cC^{-1}-\gamma) M^{(N+1)}(j, z)\label{eq:AAA}\\
 &&-\beta M^{(-N+1)}(j, z) + (\beta\cI^{(-N+1)}\cC^{-1}-\gamma) M^{(-N-1)}(j, z)  \Big)+i(\alpha  z^2 -\frac{\beta}{ z^2})\1\;.\nonumber
 \end{eqnarray}
 The last term in \eqref{eq:AAA} is irrelevant in \eqref{ZC_alg2} but allows us to obtain that $A(j, z)$ be traceless.
 By using the explicit forms of the matrices $M^{(n)}(j, z)$ given by \eqref{eq:M1}-\eqref{eq:M4}, we obtain 
 \begin{eqnarray}
 A(j, z)&=& i\omega( z) \sigma_3+ i\begin{pmatrix}
               -\beta r_j q_{j-1} -\alpha q_j r_{j-1} & 2\alpha  z q_j -\frac{2\beta}{ z} q_{j-1}\\
             2 \alpha  z r_{j-1} -\frac{2\beta}{ z} r_j & \beta r_j q_{j-1} +\alpha q_j r_{j-1}
             \end{pmatrix}\;,
      \label{eqA2}
\end{eqnarray}
where $\sigma_3$ is the Pauli matrix and
\begin{equation}
 \omega( z)=\alpha z^2+\gamma+\frac{\beta}{ z^2}\;.\label{eq:omega}
\end{equation}
This concludes our derivation of the time Lax matrix $A(j, z)$ given in \cite{XF}. Equation \eqref{ZC_alg2} now yields the zero curvature representation of \eqref{eq_Hamilton} 
 \begin{equation}\label{eq:zcper}
  \partial_{t}\ell(j, z)=A(j+1, z)\ell(j, z) - \ell(j, z) A(j, z)\;.
 \end{equation}
 In other words, the pair $(\ell(j, z), A(j, z))$ is a Lax pair for the system under consideration.   

 \subsection{B\"acklund transformation \label{sec:bap}}
 
 It is well-known that the existence of the Lax pair  $(\ell(j,z), A(j, z))$ satisfying the zero curvature equation allows one to write a consistent auxiliary problem for the $2\times 2$ matrix $\phi(j,t,z)$:
\bea
&&\phi(j+1,t,z)=\ell(j,z)\,\phi(j,t,z)\,,\label{eq:auxp1}\\
&&\partial_t \phi(j,t,z)=A(j,z)\,\phi(j,t,z)\,.\label{eq:auxp2}
\eea

For later purposes, it is important to notice that two solutions of the equations of motion can be associated 
by a B\"acklund transformation. Such a transformation has a representation at the level of the Lax pair (gauge transformation) and 
at the level of the wavefunction (Darboux transformation). For convenience, we will stick to the name B\"acklund transformation for 
all these different aspects of these transformations. 
Suppose that we have two solutions $q_j,r_j$  and $\tilde q_j,\tilde r_j$ such that there exists a matrix $B(j,t,z)$ satisfying   
\begin{equation}
B(j,t,z)\widetilde \phi(j,t,z)=\phi(j,t,z)\;,\label{eq:bff}
\end{equation}
where $\widetilde \phi(j,t,z)$ stands for the wavefunction of \eqref{eq:auxp1}-\eqref{eq:auxp2} associated with the fields $\tilde q_j,\tilde r_j$.
The matrix $B(j,t,z)$ realises a B\"acklund transformation.
The consistency of relation \eqref{eq:bff} with the auxiliary problems \eqref{eq:auxp1}-\eqref{eq:auxp2} for $ \phi(j,t,z)$ and $\widetilde \phi(j,t,z)$ leads to the following equations
for $B(j,t,z)$
 \begin{eqnarray}
 && B(j+1,t,z) \tilde \ell(j,z) =\ell(j,z) B(j,t, z)\;, \label{eq:BAper1} \\
 && \partial_t B(j,t,z) = A(j,z) B(j,t,z)- B(j,t,z) \widetilde A(j,z)\;. \label{eq:BAper2}
\end{eqnarray}
 It is easy to see that the system of the above equations is equivalent to equation \eqref{eq:BAper1} for all $j$ and equation \eqref{eq:BAper2} only for 
 one given position $j_0$, since $(\ell(j,z), A(j, z))$ and $(\tilde\ell(j,z), \widetilde A(j, z))$ satisfy the zero curvature condition.
 Then, to obtain the admissible B\"acklund transformation, we must first solve equation \eqref{eq:BAper1}. 
 The following Lemma gives one such solution that will be used below. We will deal with \eqref{eq:BAper2} in Section \ref{Aux}.
\begin{lem}\label{lem:bac}
 The matrix
 \begin{eqnarray}
  B(j,t, z)&=&\begin{pmatrix}
   z f^{1}_j +\frac { g^{1}_j }{ z}& 
  f^{1}_j\widetilde q_j -f_j^{2} q_j\\ 
-g_j^{1} r_j+g^{2}_j\widetilde  r_j&
   z f^{2}_j +\frac{g^{2}_j}{ z}
           \end{pmatrix} \nonumber \\
& +&\frac{x_j^{1}}{ z^2}\begin{pmatrix}\frac{1}{ z}&-\widetilde q_{j-1}\\
                                   -r_j- z^2 r_{j+1}(1- q_jr_j)&  z  r_j \widetilde q_{j-1}
                                  \end{pmatrix}
 +\frac{x_j^{2}}{ z^2}\begin{pmatrix}  z  \widetilde r_j q_{j-1} &  q_{j-1}\\
                                   \widetilde r_j + z^2 \widetilde r_{j+1}(1- \widetilde q_j \widetilde r_j)&\frac{1}{ z}
                                  \end{pmatrix}\nonumber\\
 &+& z^2y_j^{2}\begin{pmatrix}
                    \frac{ q_j \widetilde r_{j-1}}{ z}&- q_j-\frac{ q_{j+1}(1-r_jq_j)}{ z^2}\\
                    -\widetilde r_{j-1} &  z
                  \end{pmatrix}
 + z^2y_j^{1}\begin{pmatrix}
                    z & \widetilde q_j+\frac{  \widetilde q_{j+1}(1- \widetilde r_j  \widetilde q_j)}{ z^2}\\
                  r_{j-1} & \frac{\widetilde q_j r_{j-1}}{ z}
                  \end{pmatrix}\;.\label{eq:Ban}
\end{eqnarray}
is a solution of relation \eqref{eq:BAper1}
if the functions  $x^{1}_j$, $x^{2}_j$, $y_j^{1}$, $y_j^{2}$, $g_j^{1}$, $g_j^{2}$, $f_j^{1}$ and $f^{2}_j$ satisfy
\begin{eqnarray}
 x^{1}_{j+1}&=&x^{1}_{j} s_j \quad , \quad y_{j+1}^{2}=y^{2}_{j} s_j\quad,\quad y^{1}_{j+1}=\frac{y^{1}_{j}}{s_j} \quad , \quad x_{j+1}^{2}=\frac{x^{2}_{j}}{s_j}\\
 f^{1}_{j+1}&=& (f^{1}_{j} +y^{1}_j (q_j r_{j-1}-  \widetilde q_{j+1}\widetilde r_{j})   ) \frac{1}{s_j} \\
 g_{j+1}^{2}&=& ( g^{2}_{j} +x^{2}_j ( q_{j-1} r_j  -\widetilde q_j\widetilde r_{j+1}  )   )\frac{1}{s_j}\\
 g_{j+1}^{1}&=&(g^{1}_{j}   +x_{j}^{1} (\widetilde q_{j-1}\widetilde r_{j}- q_{j} r_{j+1}) )  s_j\\
 f^{2}_{j+1}&=&(f^{2}_{j}  +y_{j}^{2} (\widetilde q_j\widetilde r_{j-1}- q_{j+1}r_{j}))s_j
\end{eqnarray}
where $s_j=\sqrt{\frac{1-{r}_j {q}_j}{1-\widetilde r_j\widetilde q_j}}$ and the fields are constrained by
\begin{eqnarray}
&&f^{1}_j\widetilde q_j -f^{2}_j{q}_j-y^{2}_{j} {q}_{j+1}(1-q_j  r_j  ) +y^{1}_{j} \widetilde{q}_{j+1}(1- \widetilde q_j \widetilde r_j ) \nonumber \\
 &=&g^{2}_j {q}_{j-1}- g^{1}_j \widetilde q_{j-1} -x_{j}^{1} \widetilde q_{j-2}(1-\widetilde q_{j-1}\widetilde  r_{j-1})+x_{j}^{2} q_{j-2}(1- q_{j-1}  r_{j-1})  \label{eq:d12b}  \\
&&g^{2}_j\widetilde r_j- g^{1}_j r_j - x^{1}_{j} r_{j+1} (1-q_j r_j  )+ x^{2}_{j} \widetilde r_{j+1} (1- \widetilde q_j\widetilde r_j )\nonumber \\
 &=& f^{1}_j {r}_{j-1}  -f_j^{2}\widetilde r_{j-1} -y_{j}^{2}  \widetilde r_{j-2} (1-\widetilde  q_{j-1} \widetilde r_{j-1}) +y_{j}^{1}  r_{j-2} (1-  q_{j-1} r_{j-1}) \label{eq:d21b}
 \end{eqnarray}
\end{lem}
\proof The proof is done by direct computation by inserting expression \eqref{eq:Ban} into \eqref{eq:BAper1} and matching the powers of the spectral parameter $z$.
\endproof
In the solution for the B\"acklund matrix given in the previous lemma, there are 8 free parameters which we can take to be 
for example $f^{1}_0$, $f^{2}_0$, $g^{1}_0$, $g^{2}_0$, $x^{1}_0$, $x^{2}_0$, $y^{1}_0$ and $y^{2}_0$. These are determined by fixing $B(0,t,z)$.
For $x^{1}_0=x^{2}_0=y^{1}_0=y^{2}_0=0$, one gets $x^{1}_j=x^{2}_j=y^{1}_j=y^{2}_j=0$ and we recover the result of \cite{BH}, 
up to slight modifications due to the different choice of normalisation of the matrices $\ell(j,z)$.

 \section{Integrable Ablowitz--Ladik model on the finite lattice}\label{AL_openBC}
 
 \subsection{Double-row transfer matrix}
 
To consider open finite chains with integrable boundary conditions, one needs to modify the single-row method of the previous section. In his seminal work, Sklyanin \cite{Sk} proposed to consider the so-called double-row transfer matrix instead of the single-row transfer matrix \eqref{single_mono}. 
It is defined by
\begin{equation}\label{eq:tr-dr}
 b( z)=\text{tr}_a\left( k^+_a( z) \mathbb{L}_a( z) \right) \quad \text{with}\quad  \mathbb{L}_a( z)=L_a( z) k_a^-( z) L_a(\tau( z))^{-1}\;,
\end{equation}
where $L_a(z)$ is defined as in \eqref{single_row}.
The matrices $k^\pm(u)$ are reflection matrices describing the boundary conditions at both ends of the finite chain and $\tau$ is a 
function of the spectral parameter that depends on the model. In our case,
\begin{equation}
 \tau( z)=\sqrt{\frac{\beta}{\alpha}} \ \frac{1}{ z}\;.
\end{equation}
Let us emphasize that $\tau$ is chosen so that $\omega(\tau( z))=\omega( z)$ where $\omega$ is given by \eqref{eq:omega} and that $\tau(\tau( z))= z$. In the historical paper \cite{Sk}, 
a simpler involution $\tau( z)=1/ z$ was used but we must generalize that construction to find integrable boundary conditions for any $\alpha$ and $\beta$. 
The matrices $k^-( z)$ and  $k^+(\tau( z))$ are required to be solutions of the reflection equation 
\begin{equation}
\label{rkk1}
r_{ab}\left(\frac{ w}{ z}\right)k_a( w) k_b( z) +   k_a( w)   \ k_b( z) r_{ab}\left(\frac{\tau( w)}{\tau( z)}\right)
-k_a( w)r_{ab}\left(\frac{\tau( w)}{ z}\right) k_b( z)-k_b( z)r_{ab}\left(\frac{ w}{\tau( z)}\right) k_a( w)=0\,.
\end{equation}
These equations ensure that $\mathbb{L}_a( z)$ satisfies the so-called dynamical reflection equation
\begin{eqnarray}
\{\mathbb{L}_a( w),\mathbb{L}_b( z)\}&=&
 r_{ab}\left(\frac{ w}{ z}\right)\mathbb{L}_a( w)\mathbb{L}_b( z) +  \mathbb{L}_a( w)\mathbb{L}_b( z) r_{ab}\left(\frac{\tau( w)}{\tau( z)}\right)\nonumber\\
&&-\mathbb{L}_a( w)r_{ab}\left(\frac{\tau( w)}{ z}\right) \mathbb{L}_b( z)-\mathbb{L}_b( z)r_{ab}\left(\frac{ w}{\tau( z)}\right) \mathbb{L}_a( w)\,,
 \end{eqnarray}
which in turn allows one to prove the the analog of the important result \eqref{PB_commute} for $b(u)$ \ie
\be
\label{PB_com_dr}
\{ b( z), b( w)\}=0\,.
\ee
We take the following solutions of \eqref{rkk1} as our reflection matrices
\begin{eqnarray}
  k^-( z)=  \begin{pmatrix}
            a  z +\frac{b}{\alpha  z} & c\left( z^2\sqrt{\frac{\alpha}{\beta}}-\frac{1}{ z^2}\sqrt{\frac{\beta}{\alpha}}\right)\\
              d\left( z^2-\frac{\beta}{ z^2\alpha}\right)&\frac{a}{ z}\sqrt{\frac{\beta}{\alpha}} +\frac{b  z}{\sqrt{\alpha\beta}}
             \end{pmatrix}
\quad,\qquad
  k^+( z)= \begin{pmatrix}
               z  &0\\
              0&\frac{1}{ z}\sqrt{\frac{\beta}{\alpha}}
             \end{pmatrix}
\end{eqnarray}
where $a,\ b,\ c$ and $d$ are arbitrary parameters. 
The matrix $k^-( z)$ is the most general solution of the reflection equation (up to an irrelevant overall scaling). As we shall see, it describes the boundary conditions at the ``origin''. 
The matrix $k^+( z)$ describes the vanishing of the fields at the site $N+1$ and is chosen for simplicity here. 
In particular, this choice allows us to consider the ``half-line'' problem with vanishing fields at infinity, simply by taking $N$ to infinity. 
Those choices are already general enough to obtain new integrable boundary conditions for the Ablowitz--Ladik model on the half-line which contain the (discrete) 
Robin boundary condition as a particular case. 

The expansion of the double-row transfer matrix $b( z)$ \eqref{eq:tr-dr} is written as follows
 \be
 b( z)=\frac{\II^{(0)}}{ z^{2N+4}}+\frac{\II^{(1)}}{ z^{2N+2}}  +\dots  
 \ee 
 From relation \eqref{PB_com_dr}, we deduce that $\{ \II^{(n)},\II^{(p)}\}=0$.
Upon inspection of the bulk terms, we define the Hamiltonian by the following combination of the previous charges:
\begin{equation}\label{eq:HH}
 \HH=-2\beta \frac{\II^{(1)}}{\II^{(0)}}-2\gamma \ln\left( \left(\frac{\alpha}{\beta}\right)^{N/2+3}\II^{(0)}\right)\;.
\end{equation}
By using the explicit form of $b( z)$, we can show that the Hamiltonian is given explicitly by
\begin{equation}
\label{eq:H}
 \HH= -2 \sum_{j=0}^{N-1} (\alpha r_jq_{j+1} +\beta q_jr_{j+1})+2\gamma \sum_{j=0}^{N} \ln(1-q_jr_j) +\mathbb{B}(q_0,r_0,q_1,r_1)
\end{equation}
with
\begin{eqnarray}
\mathbb{B}(q_0,r_0,q_1,r_1)&=&-2 \frac{(1-q_0r_0)(b+\alpha d q_1-\beta c r_1)}{a+d q_0-cr_0}-2\gamma \ln(a+d q_0-cr_0)\;.
\end{eqnarray}
Comparing with \eqref{AL_Ham}, we see that the effect of the boundary is contained in $\mathbb{B}$ and involves the two neighbouring sites $j=0,1$ of the origin. 
From now on, the time evolution is associated to this Hamiltonian $\HH$ by 
\begin{equation}
 \partial_{t}\ \boldsymbol{\cdot}=\{ \HH ,\ \boldsymbol{\cdot} \}\,.
\end{equation}
Then, the Hamilton's equations of motion can be computed using \eqref{PB_fields}. Explicitely, on the left boundary they read
\begin{alignat}{3}
&\dot{q}_0=2i\left(\alpha q_1 +\gamma q_0 -\alpha q_0r_0q_1 +\frac{1-q_0r_0}{a+dq_0-cr_0}
       \left(\frac{(c-aq_0-dq_0^2)(b+\alpha dq_1-\beta cr_1)}{a+dq_0-cr_0} -\gamma c  \right)  \right)\label{eq:eomd}\\
&\dot{r}_0=-2i\left(\beta r_1+\gamma r_0-\beta q_0r_0r_1 -\frac{1-q_0r_0}{a+dq_0-cr_0}
          \left(\frac{(d +ar_0- cr_0^2)(b+\alpha dq_1-\beta cr_1)}{a+dq_0-cr_0} - \gamma d\right)\right)\label{eq:eomdr}\\
&\dot{q}_1=2i\left( \alpha q_2+\gamma q_1+\beta q_0 -q_1r_1 (\alpha q_2+\beta q_0)-\frac{\beta c(1-q_0r_0)(1-q_1r_1)}{a+dq_0-cr_0}  \right)\\
&\dot{r}_1=-2i\left( \beta r_2+\gamma r_1+\alpha r_0 -q_1r_1(\alpha r_0+\beta r_2)+\frac{\alpha d(1-q_0r_0)(1-q_1r_1)}{a+dq_0-cr_0}\right)
\label{eq:eomf1}
\end{alignat}
while in the bulk they are given by, for $j=2,3\dots,N-1$,
\begin{alignat}{3}
&\dot{q}_j=2i\big(\alpha q_{j+1}+\gamma q_j+\beta q_{j-1} -q_jr_j(\alpha q_{j+1}+\beta q_{j-1})\big)\\
&\dot{r}_j=-2i\big(\beta r_{j+1}+\gamma r_j+\alpha r_{j-1} -q_jr_j(\alpha r_{j-1}+\beta r_{j+1})\big) \label{eq:eomf2}
\end{alignat}
and on the right boundary, we have
\begin{alignat}{3}
&\dot{q}_N=2i\big(\gamma q_N+\beta q_{N-1} -\beta q_Nr_N q_{N-1}\big)\\
&\dot{r}_N=-2i\big(\gamma r_N+\alpha r_{N-1} -\alpha q_Nr_N r_{N-1}\big) \label{eq:eomf3}.
\end{alignat}
The equations of motion on the finite lattice come from the Hamiltonian \eqref{eq:HH} which was extracted from the transfer matrix $b(z)$. This  gives a proof that they are integrable in the sense of Liouville. We now turn to the derivation of the time Lax matrix producing these on the finite open lattice. 

\subsection{Double-row Lax pair}\label{double_row_LP}

Recall that using the $r$-matrix approach, we can derive the time Lax matrices for all the time flows associated to the Lax matrix $\ell(j,z)$
via its single-row transfer matrix, as well as the corresponding zero curvature equations reproducing Hamilton's equations of motion. The same holds true for models with boundaries but, of course, one has to modify formulas \eqref{Mbar} and \eqref{M} to take into account the presence of integrable boundaries, as dictated by the double-row transfer matrix.
Following the construction of \cite{AD1}, we define, for $j=0,1,\dots,N+1$, 
\begin{eqnarray}\label{eq:Mad}
\MM_b(j, w, z)&=&tr_a\left(\ k^+_a( w) \ L_a(N,j, w)\  r_{ab}\left(\frac{ w}{ z}\right)\ L_a(j-1,0, w) \ k^-_a( w)\ L_a(\tau( w))^{-1}\ \right)\nonumber\\
&-&tr_a\left(\ k^+_a( w)  \ L_a( w)\  \ k^-_a( w)\ L_a(j-1,0,\tau( w))^{-1}\ r_{ab}\left(\frac{\tau( w)}{ z}\right)\ L_a(N,j,\tau( w) )^{-1}\ \right)\;.
\end{eqnarray}
These matrices satisfy the following property \cite{AD1}
\begin{equation}\label{eq:bM}
\{ b( w) , \ell_a(j, z) \} = \MM_a(j+1, w, z)\ell_a(j, z) - \ell_a(j, z) \MM_a(j, w, z)\;.
\end{equation}
Note that the proof of \cite{AD1} must be adapted here to take in account our involution $\tau$. In \cite{ACC}, it was shown that $\MM(j, w, z)$ satisfies two other equations which are of crucial importance for describing integrable boundary conditions in a zero curvature representation
\begin{eqnarray}
&&\MM_a(0, w, z) k_a^-( z) - k^-_a( z) \MM_a(0, w,\tau( z))=0 \label{eq:Mkm} \\
&&\MM_a(N+1, w,\tau( z)) k_a^+( z) - k^+_a( z) \MM_a(N+1, w, z)=0\label{eq:Mkp} 
\end{eqnarray}

Writing the expansion of  $\MM(j, w, z)$ as 
\be
\frac{1}{ w^{2N+4}}\MM^{(0)}(j, z)+\frac{1}{ w^{2N+2}}\MM^{(1)}(j, z)+\dots
\ee
and expanding \eqref{eq:bM}, \eqref{eq:Mkm} and \eqref{eq:Mkp}  w.r.t. $ w$ in accordance with expression \eqref{eq:HH} for $\HH$  in terms of the expansion of $b(u)$, we can derive the following 
\begin{eqnarray}
\label{bd_eqs_motion}
&&\partial_{t}\ \ell(j, z)=\{\HH,\ell(j, z)\}=\Ad(j+1, z)\ell(j, z) - \ell(j, z) \Ad(j, z)\\
\label{eq:ZCB1}&&  \partial_{t}\ k^-( z)=0=\Ad(0, z)\ k^-( z) - k^-( z)\ \Ad(0,\tau( z))\\
\label{eq:ZCB2}&&   \partial_{t}\ k^+( z)=0=\Ad(N+1,\tau( z))\ k^+( z) - k^+( z)\ \Ad(N+1, z)
\end{eqnarray}	
where 
\begin{equation}
\Ad(j, z)=-2\beta\frac{\MM^{(1)}(j, z)}{\II^{(0)}}+2\beta\frac{\II^{(1)}\ \MM^{(0)}(j, z)}{(\II^{(0)})^2}-2\gamma \frac{\MM^{(0)}(j, z)}{\II^{(0)}} +i\omega( z)\;,\label{eq:AA}
\end{equation}
where $\omega( z)$ is given by \eqref{eq:omega}. 
The zero curvature equations \eqref{bd_eqs_motion}-\eqref{eq:ZCB2} are invariant by adding a term to $\Ad(j, z)$ which is proportional to the identity, 
is independent of $j$ and is invariant under $ z\to \tau( z)$. We use this freedom in \eqref{eq:AA} to add $i\omega( z)$ and make $\Ad(j, z)$ traceless.
Then $(\ell(j, z),\Ad(j, z))$ is an adequate Lax pair associated with the AL model with integrable boundary conditions determined by the matrices $k^\pm(z)$. The status of relations 
\eqref{eq:ZCB1}-\eqref{eq:ZCB2} is discussed in detail in \cite{ACC}.

Upon performing the explicit computations of the matrices $\MM^{(0)}(j, z)$, $\MM^{(1)}(j, z)$, we find that the matrices $\Ad(j,z)$ for $j=2,\dots,N$ have exactly the same form as the ones with periodic boundary condition. Namely, one gets
\begin{eqnarray}
\Ad(j, z)= A(j, z)=i \begin{pmatrix}
\omega( z)-\alpha q_j r_{j-1}-\beta q_{j-1}r_j &2\alpha q_j  z-2\beta q_{j-1}/ z\\
2\alpha r_{j-1}  z- 2\beta r_j/ z & -\omega( z)+\alpha q_j r_{j-1}+\beta q_{j-1}r_j
\end{pmatrix} \;.\label{eq:Aj}
\end{eqnarray}
However, the matrices $\Ad(0,z)$ and $\Ad(1, z)$ are different
\begin{eqnarray}
\Ad(1, z)=A(1, z)
+i\frac{1-q_0r_0}{dq_0-cr_0+a} \begin{pmatrix}\beta c r_1-\alpha dq_1&\beta c/ z\\
\alpha d  z & \alpha d q_1 -\beta c r_1
\end{pmatrix} 
\end{eqnarray}
and
\begin{eqnarray}
\Ad(0, z)&=&\frac{i}{a+d q_0-cr_0} \left\{
\left(\omega(  z)- \frac{(1-q_0r_0)(b+\alpha d q_1-\beta c r_1)}{a+d q_0-cr_0} \right)
\begin{pmatrix}
a&2c/ z\\
2d  z & -a
\end{pmatrix}\right.\label{eq:Ad1} \\
&&\left. +b
\begin{pmatrix}
1+q_0r_0 & 2q_0/ z\\
-2r_0 z& -1-q_0r_0
\end{pmatrix}
-\begin{pmatrix}
(cr_0+dq_0)(\alpha z^2-\beta/ z^2) & 2\alpha (c-aq_0+cq_0r_0) z\\
2\beta (d+ar_0+dq_0r_0) / z  &(cr_0+dq_0)(\beta/ z^2-\alpha z^2)
\end{pmatrix} \right\}\;.\nonumber
\end{eqnarray}
Similarly, the matrix $\Ad(N+1, z)$ is also different from the bulk one and is given by
\begin{eqnarray}
\Ad(N+1, z)&=&i\begin{pmatrix}
\omega( z) & -2\beta q_{N}/ z\\
2\alpha r_{N}  z  &-\omega( z)
\end{pmatrix}\;.
\end{eqnarray}        
This is the signature on the Lax matrices of the presence of boundary conditions.
These explicit expressions can of course be used to check directly the validity of \eqref{eq:ZCB1}-\eqref{eq:ZCB2} and the 
equivalence between \eqref{bd_eqs_motion} and \eqref{eq:eomd}-\eqref{eq:eomf3}.

 \section{Ablowitz--Ladik model on the half-infinite lattice 	with time-dependent integrable boundary conditions}\label{ALboundary}
 
\subsection{Intrinsic vs extrinsic picture for boundary conditions \label{sec:cv}}

\subsubsection{Illustrating the idea on Robin boundary conditions}

The reader familiar with integrable chain models on a finite interval will be perfectly content with the {\it intrinsic} representation of 
the boundary conditions at the end of the interval {\it within} the equations of motion as in \eqref{eq:eomd}-\eqref{eq:eomf3}.
However, the reader who is more familiar with integrable PDEs on the finite interval (or half-line) might be more used to an {\it extrinsic} 
representation of the problem in the form of a bulk equation of motion valid {\bf for all} values of the space coordinates supplemented by a 
condition on the field (and spatial derivatives) at the coordinate of the boundary. 

To clarify what we mean, let us first consider \eqref{eq:eomd}-\eqref{eq:eomf3} in the case $c=d=0$, which corresponding the (discrete) Robin condition on the left boundary as we will see. The intrinsic picture is given by eqs \eqref{eq:eomd}-\eqref{eq:eomf3} which boil down to
the equations in the bulk, now valid for $j=1,\dots,N$,
\bea
\label{Robin1}&&\dot{q}_j=2i(\alpha q_{j+1}+\gamma q_j+\beta q_{j-1} -q_jr_j(\beta q_{j-1}+\alpha q_{j+1}))\\
\label{Robin2}&&\dot{r}_j=-2i(\beta r_{j+1}+\gamma r_j+\alpha r_{j-1} -q_jr_j(\alpha r_{j-1}+\beta r_{j+1})).
\eea
the equations at the left boundary,
\bea
\label{Robin3}&&\dot{q}_0=2i(\alpha q_1 + \gamma q_0 -\alpha q_0r_0q_1-\frac{b}{a}(1-q_0r_0)q_0) \\
\label{Robin4}&&\dot{r}_0=-2i(\beta r_1+\gamma r_0-\beta q_0r_0r_1 -\frac{b}{a}(1-q_0r_0)r_0)
\eea
and the ones at the right boundary,
\begin{alignat}{3}
&\dot{q}_N=2i\big(\gamma q_N+\beta q_{N-1} -\beta q_Nr_N q_{N-1}\big)\\
&\dot{r}_N=-2i\big(\gamma r_N+\alpha r_{N-1} -\alpha q_Nr_N r_{N-1}\big).
\end{alignat}
We see that the equations at $j=0$ and at $j=N$ are different from the naive continuation of the bulk equations for $j=1,\dots,N-1$ to $j=0$ or $j=N$. 
In the extrinsic picture, the idea is to enforce this continuation and to think of the boundaries as sitting at $j=-1$ and $j=N+1$. To produce a system of equation equivalent to the one above, we introduce the value of the fields $q_{-1}$, $r_{-1}$, $q_{N+1}$ and $r_{N+1}$ at these sites such that the bulk equations of motion are the same {\it for all} $j= 0,\dots,N$
\bea
&&\dot{q}_j=2i(\alpha q_{j+1}+\gamma q_j+\beta q_{j-1} -q_jr_j(\beta q_{j-1}+\alpha q_{j+1})) \\
&&\dot{r}_j=-2i(\beta r_{j+1}+\gamma r_j+\alpha r_{j-1} -q_jr_j(\alpha r_{j-1}+\beta r_{j+1})).
\eea
Then, at $j=0$, the equivalence with the intrinsic picture is restored when imposing the boundary conditions 
\be
\beta q_{-1}+\frac{b}{a}q_0=0\,,~~\alpha r_{-1}+\frac{b}{a}r_0=0\,.\label{eq:ro1}
\ee
In the discrete NLS case $\alpha=\beta=1/2$, these are known to be the discrete analog of the Robin boundary conditions, as studied e.g. in \cite{BH}. At $j=N$, the equivalence with the intrinsic picture is obtained by setting
\be
q_{N+1}=0\,,~~r_{N+1}=0\,,
\ee
which are Dirichlet boundary conditions. 

In the rest of the paper, we keep these conditions on the right boundary and send $N\to\infty$ to consider the system on the half-infinite lattice with zero boundary conditions at infinity. We also restore $c,d\neq 0$ in order to consider our general boundary conditions in the extrinsic picture.

\subsubsection{From intrinsic to extrinsic for our general boundary conditions: emergence of time-dependent boundary conditions}

Motivated by this discussion, we would like to interpret our more general intrinsic equations \eqref{eq:eomd}-\eqref{eq:eomf3} from the extrinsic point of view, in the case of 
arbitrary parameters $a,\ b,\ c,\ d$. A difficulty arises since the intrinsic equations of motion are modified both on site $j=0$ and $j=1$ 
so that an interpretation via a boundary sitting at $j=-1$ together with a condition relating the values of the fields at $j=-1$ and $j=0$ is not clear. 
To circumvent this problem, it is convenient to perform a change of variables.
Let us define the new fields 
\begin{equation}
 Q_0=q_0+\frac{c(q_0r_0-1)}{a+dq_0-cr_0}\,,~~ R_0=r_0-\frac{d(q_0r_0-1)}{a+dq_0-cr_0}
 \,,~~Q_j=q_j\,,~~R_j=r_j\,,~~j\ge 1\;.\label{eq:cgvar}
\end{equation}
The inversion of the change of variables \eqref{eq:cgvar} gives
\begin{eqnarray}
q_0=Q_0-\frac{1}{2d}\left(  a \pm \sqrt{4cd(1-Q_0R_0)+a^2}\right)\quad\text{and}\qquad 
r_0=R_0+\frac{1}{2c}\left(  a \pm \sqrt{4cd(1-Q_0R_0)+a^2}\right)\;.
\end{eqnarray}
In these new variables, the equations of motion for $j\ge 1$ read
\begin{alignat}{3}
&\dot{Q}_j=2i\big(\alpha Q_{j+1}+\gamma Q_j+\beta Q_{j-1} -Q_jR_j(\alpha Q_{j+1}+\beta Q_{j-1})\big)\label{eq:eom1}\\
&\dot{R}_j=-2i\big(\beta R_{j+1}+\gamma R_j+\alpha R_{j-1} -Q_jR_j(\alpha R_{j-1}+\beta R_{j+1})\big)\;.\label{eq:eom2}
\end{alignat}
In particular, the equation of motion at the site $j=1$ has now the same form as the other ones in the bulk.
Therefore, after this transformation, all the effect of the boundary is carried by the equations of motion at site $0$, similarly to the 
Robin case \eqref{Robin1}-\eqref{Robin4}. It remains to go over the extrinsic picture by introducing the values of the fields $Q_{-1}$ and $R_{-1}$
in such a way that the bulk equations of motion are formally the same for all $j\ge 0$ and are supplemented by boundary conditions involving 
$Q_{-1}$ and $R_{-1}$ and neighbouring sites. 

The time derivative of $Q_0$ and $R_0$ are easily computed 
\begin{eqnarray}
 \dot Q_0&=&\dot q_0 +\frac{c}{(a+dq_0-cr_0)^2}\left( (d+ar_0-cr_0^2)\dot q_0 -(c-aq_0-dq_0^2) \dot r_0   \right)\;,\\
 \dot R_0&=&\dot r_0 -\frac{d}{(a+dq_0-cr_0)^2}\left( (d+ar_0-cr_0^2)\dot q_0 -(c-aq_0-dq_0^2) \dot r_0   \right)\;.
\end{eqnarray}
Then, by using the equations of motion  \eqref{eq:eomd} and \eqref{eq:eomdr}, the previous relations can be written
\begin{eqnarray}
 \dot Q_0&=&2i\left( \alpha Q_1 +\gamma Q_0 -\alpha Q_0R_0 Q_1  +\frac{(1-q_0r_0) \big( \alpha cd q_1(1-q_0r_0)+ b(c-aq_0-dq_0^2)  \big) }{(a+dq_0-cr_0)^2}  \right) \\
 \dot R_0&=&-2i\left( \beta R_1 +\gamma R_0 -\beta Q_0R_0 R_1  +\frac{(1-q_0r_0) \big(  \beta cd r_1(1-q_0r_0) -b(d+ar_0-cr_0^2)    \big) }{(a+dq_0-cr_0)^2}  \right)
\end{eqnarray}
In view of this, the equations of motion at the site $0$ for $Q_0$ and $R_0$ can be written as the equation of motion in the bulk \eqref{eq:eom1}-\eqref{eq:eom2} continued to $j=0$, \ie:
\begin{eqnarray}
 \dot Q_0&=&2i\left( \alpha Q_1 +\gamma Q_0 +\beta Q_{-1} - Q_0R_0 (\alpha Q_1+\beta Q_{-1})  \right)\\
 \dot R_0&=&-2i\left( \beta R_1 +\gamma R_0 +\alpha R_{-1} - Q_0R_0(\beta  R_1 +\alpha R_{-1})  \right)
\end{eqnarray}
provided $Q_{-1}$ and $R_{-1}$ satisfy the following conditions
\begin{eqnarray}
 Q_{-1}&=&\frac{\alpha}{\beta}Q_1+\frac{ (a\alpha Q_1+bQ_0)\left(a\pm \sqrt{4cd(1-Q_0R_0)+a^2}\right)}{2cd\beta (1-Q_0R_0)}\label{eq:bh1}\\
 R_{-1}&=&\frac{\beta}{\alpha}R_1+\frac{ (a\beta R_1+bR_0)\left(a\pm \sqrt{4cd(1-Q_0R_0)+a^2}\right)}{2cd\alpha (1-Q_0R_0)}\;.\label{eq:bh2}
\end{eqnarray}
We can summarize the previous discussion in the following proposition.
\begin{prop} The Hamilton equations of motion obtained from the Liouville integrable Hamiltonian \eqref{eq:H} are equivalent to 
the following equations in the extrinsic picture, for $j\ge 0$,
 \begin{alignat}{3}
&\dot{Q}_j=2i\big(\alpha Q_{j+1}+\gamma Q_j+\beta Q_{j-1} -Q_jR_j(\alpha Q_{j+1}+\beta Q_{j-1})\big)\;,\label{eq:eomb1}\\
&\dot{R}_j=-2i\big(\beta R_{j+1}+\gamma R_j+\alpha R_{j-1} -Q_jR_j(\alpha R_{j-1}+\beta R_{j+1})\big)\;, \label{eq:eomb2}
\end{alignat}
together with the boundary conditions \eqref{eq:bh1} and \eqref{eq:bh2}, under the change of variables \eqref{eq:cgvar}.
\end{prop}
Let us remark that the boundary conditions
\eqref{eq:bh1}-\eqref{eq:bh2} depends on three sites -1,0 and 1. However, by using equations of motion \eqref{eq:eomb1}-\eqref{eq:eomb2} at site $0$, 
we can eliminate $Q_1$ and $R_1$ to get boundary conditions depending only on the two sites -1 and 0. The time
derivative of the fields at the site $0$ then appears. 
Namely, the boundary conditions \eqref{eq:bh1}-\eqref{eq:bh2} become
\begin{eqnarray}
 Q_{-1}&=&\frac{icd\dot Q_0+\Big(2\gamma c d +ab\mp b\sqrt{4cd(1-Q_0R_0)+a^2}\Big) Q_0}{2cd\beta (1-Q_0R_0)}\;,\label{eq:bh1tp}\\
 R_{-1}&=&\frac{-icd\dot R_0+\Big(2\gamma c d +ab\mp b\sqrt{4cd(1-Q_0R_0)+a^2}\Big) R_0}{2cd\alpha (1-Q_0R_0)}\;.\label{eq:bh2tp}
\end{eqnarray}
Due to the presence of the time derivative in the boundary conditions, we call them time-dependent boundary conditions.
This terminology has in fact a deeper root as will become clear later on: the reflection matrix describing them is time-dependent. 
This has non-trivial consequences on the implementation of the so-called nonlinear mirror image method which we address below.

\subsection{Reductions}

The reduction of the previous equations of motion to get DNLS or DMKdV leads to some constraints on the boundary parameters.
For the DNLS (\textit{i.e.} $\alpha=\beta=1/2$, $\gamma=-1$ and $R_j^*=\nu Q_j$), the extrinsic equations of motion becomes 
\begin{equation}
\dot{Q}_j=i\big( Q_{j+1}-2 Q_j+Q_{j-1} -\nu |Q_j|^2( Q_{j+1}+ Q_{j-1})\big)\;,\qquad\text{for }j\ge 0
\end{equation}
with the boundary conditions
\begin{equation}
Q_{-1}= Q_1+\frac{ (a Q_1+2b Q_0)\left(a\pm \sqrt{4cd(1-\nu|Q_0|^2)+a^2}\right)}{2cd (1-\nu|Q_0|^2)}\;.
\end{equation}
The parameters of the boundary satisfy $a,b\in \RR$, $c=-\nu d^*$.

For the DMKdV (\textit{i.e.} $\alpha=-\beta=i/2$, $\gamma=0$, $R_j=\nu Q_j$ and $R_j\in \RR$), the extrinsic equations of motion becomes 
\begin{equation}
 \dot{Q}_j=Q_{j-1}- Q_{j+1}  +\nu Q_j^2( Q_{j+1}- Q_{j-1})\;,\label{eq:dkdvre}
\end{equation}
with the boundary conditions
\begin{equation}
Q_{-1}=- Q_1-\frac{ a Q_1\left(a\pm \sqrt{4cd(1-\nu|Q_0|^2)+a^2}\right)}{2cd (1-\nu|Q_0|^2)}\;,
\end{equation}
The parameters $a,b,c,d$ of the boundary should all be real and satisfy $c=-\nu d$, $b=0$.

\section{Lax pair and zero curvature equations in the extrinsic picture}\label{LP}

In view of the previous discussion, we need to establish the effect of going from intrinsic to extrinsic picture on the results of Section \ref{double_row_LP}.

\subsection{Change of variable, gauge transformation and extrinsic picture in the Lax pair presentation}

By using the change of variable \eqref{eq:cgvar}, we get new expressions for the matrices $\Ad(j, z)$ for $j\ge 1$,
\begin{eqnarray}
 \Ad(j, z)&=&
 i \begin{pmatrix}
       \omega( z)-\alpha Q_j R_{j-1}-\beta Q_{j-1}R_j &2\alpha Q_j  z-2\beta Q_{j-1}/ z\\
        2\alpha R_{j-1}  z- 2\beta R_j/ z & -\omega( z)+\alpha Q_j R_{j-1}+\beta Q_{j-1}R_j
        \end{pmatrix} \;.\label{eq:AjQ}
 \end{eqnarray}
We see in particular that the Lax matrix $\Ad(1, z)$ on site $j=1$ takes the same form as the bulk ones $\Ad(j, z)$, $j\ge 2$. 
The same feature appears already in Section \ref{sec:cv} where the equations of motion on the site $j=1$ becomes similar to the 
ones of the bulk after the change of variable. However, even after the change of variables, the matrix $\Ad(0, z)$ still has a different structure from the bulk ones. This is not compatible with the spirit of the extrinsic picture and we have seen that introducing fields
 $Q_{-1}$ and $R_{-1}$ satisfying appropriate boundary conditions (see \eqref{eq:bh1}-\eqref{eq:bh2}) allowed us to have equations of motion with the same form for all $j\ge 0$. Therefore, we would like to introduce the same fields such that the Lax matrix $\Ad(0, z)$ is written as in the bulk, \ie
\begin{eqnarray}
 \Ad(0, z)&=&
 i \begin{pmatrix}
       \omega( z)-\alpha Q_0 R_{-1}-\beta Q_{-1}R_0 &2\alpha Q_0  z-2\beta Q_{-1}/ z\\
        2\alpha R_{-1}  z- 2\beta R_0/ z & -\omega( z)+\alpha Q_0 R_{-1}+\beta Q_{-1}R_0
        \end{pmatrix} \;,
\label{eq:AjQ0}
 \end{eqnarray}
 and such that the equation 
 \begin{equation}
 \Ad(0, z)\ k^-( z) - k^-( z)\ \Ad(0,\tau( z))=0 \label{eq:rob}
 \end{equation}
 is equivalent to the boundary conditions \eqref{eq:bh1}-\eqref{eq:bh2}. However, for the generic boundary conditions associated to $k^-(z)$, this procedure fails 
 since $\Ad(0, z)$ given by \eqref{eq:Ad1} cannot be written in the form \eqref{eq:AjQ0}. This is readily seen from the difference in the dependence on the spectral parameter $ z$ which cannot be accommodated by a constraint involving only fields. This is a feature of our new boundary conditions. Indeed, for the particular choice of the Robin boundary conditions ($c=d=0$), 
 the passage to the extrinsic picture actually works and equation \eqref{eq:rob} with $\Ad(0, z)$ given by \eqref{eq:AjQ0} is equivalent to the Robin conditions \eqref{eq:ro1}.
 To overcome this problem for generic boundary condition, we consider a gauge transformation $G( z)$ concentrated at site $0$ and defined by
\begin{eqnarray}
\cL(0, z)=\ell(0, z) G( z)^{-1}\quad,\qquad {\cL}(j, z)=\ell(j, z)\quad j\ge 1\,,
\end{eqnarray}
with, for $j\ge 0$
\begin{eqnarray}
{\cL}(j, z)&=&\frac{1}{\sqrt{1-Q_jR_j}}\begin{pmatrix}
 z &Q_j\\
R_j &\frac{1}{ z}
\end{pmatrix}\;.
\end{eqnarray}
We find that the gauge transformation is given in terms of the fields $q_0$ and $r_0$ by
\begin{eqnarray}
 G( z)&=&\frac{1}{\sqrt{ (a+dq_0)(a-cr_0)+cd  }}\begin{pmatrix}
    a+dq_0& \frac{c}{ z}\\
    -d z& a-cr_0
   \end{pmatrix}\;.
\end{eqnarray}
Let us emphasize that since the gauge transformation depends on the fields $q_0$ and $r_0$, its time derivative does not vanish.
By injecting the gauge transformation for $\cL(j, z)$ into the zero curvature equation \eqref{bd_eqs_motion}, one gets that the Lax matrices $\Ad(j, z)$ must be transformed to
\begin{eqnarray}
&&\cA(0, z)=\dot G( z) G( z)^{-1} + G( z)\Ad(0, z) G( z)^{-1}\,,~~\cA(j, z)=\Ad(j, z)\,\ j\ge 1\;.
\end{eqnarray}
The remarkable result is that (after some tedious computations) we can show that $\cA(0, z)$ can indeed be written as a Lax matrix of the same form as the bulk ones and takes the form \eqref{eq:AjQ0}
where $Q_{-1}$ and $R_{-1}$ satisfies \eqref{eq:bh1} and \eqref{eq:bh2}, as desired. Therefore, we have succeeded in performing the transition from intrinsic to extrinsic picture consistently at the level of the Lax pair:
\be
(\ell(j,z),\Ad(j,z))\mapsto (\ell(j,z),\cA(j,z))\,.
\ee

Perhaps the most important feature of this transition is the effect of the gauge transformation on the boundary matrix $k^-(z)$
\begin{equation}
k^-(z)\mapsto K^-( z)=G( z) \ k^-( z)\ G(\tau( z))^{-1}\,.
\end{equation}
It is explicitly given by
\begin{equation}
\label{TD_Kmat}
  K^-( z)= \begin{pmatrix}
                        a z+\frac{b}{\alpha z}& 0\\
                        0& \frac{a\sqrt{\beta} }{\sqrt{\alpha} z}+\frac{b z}{\sqrt{\alpha\beta}}
                       \end{pmatrix}
+\frac{(\frac{\beta}{\alpha z^2}-  z^2)\left(a\pm \sqrt{4cd(1-Q_0R_0)+a^2}\right)}{2(1-Q_0R_0)}
\begin{pmatrix}
 \frac{1}{  z}& Q_0\sqrt\frac{\alpha}{\beta}\\
 -R_0&-  z\sqrt\frac{\alpha}{\beta}
\end{pmatrix}\;.
\end{equation}
In turn this has a nontrivial impact on the boundary zero curvature equation \eqref{eq:ZCB1} which now reads 
\begin{equation}
\label{time_dep_BC}
  \partial_{t}\  K^-( z)=\cA(0, z)K^-( z)-K^-( z)\ \cA(0,\tau( z))\,.
\end{equation}
This equation is the main reason for calling these boundary conditions {\it time-dependent}: \eqref{time_dep_BC} is the time-dependent generalisation of \eqref{eq:ZCB1}. Of course, it is not the first time that this equation appears in relation to integrable boundary conditions. However, as explained in \cite{ACC} and as illustrated on the AL model here, it is the first time that it is directly linked to the Hamiltonian approach 
and that it appears as a necessity to give an extrinsic picture of boundary conditions corresponding to the most general solutions of the 
reflection equation.
Finally, we can now give the extrinsic form of the equations of motion in the zero curvature representation.
\begin{prop}\label{pro:ZC}
 The equations of motion given by relations \eqref{eq:eom1}-\eqref{eq:eom2} 
 for $j\ge 0$ and the boundary conditions \eqref{eq:bh1} and \eqref{eq:bh2}
 are equivalent to 
 \begin{eqnarray}
\label{bd_eqs_motioncL}
&&\partial_{t}\ \cL(j, z)=\cA(j+1, z)\cL(j, z) - \cL(j, z) \cA(j, z)\;,\ \ \text{for } j\ge 0 \\
\label{eq:ZCB1K}&&  \partial_{t}\ K^-( z)=\cA(0, z)K^-( z)-K^-( z)\ \cA(0,\tau( z))\;.
\end{eqnarray}	
\end{prop}
\proof The procedure explained previously provides the proof that the equations of motion imply the zero curvature representation. 
The implication in the other way is proven by direct computation.  \endproof

In other words, the boundary matrix $K^-( z)$ is now dynamical and describes time-dependent boundary conditions, even though the original 
boundary matrix $k^-(z)$ was non-dynamical. We wish to stress at this point 
that this observation was one of the main motivation behind \cite{ACC}. We see that if one want to consider the most general solution 
of the {\bf non}-dynamical reflection equation \eqref{rkk1} and have an {\bf extrinsic} interpretation of the boundary conditions, one is
naturally led to consider the dynamical setting. It is at the basis of the long-standing discrepancy between the classical $r$-matrix approach to boundary conditions and the zero curvature approach. This was resolved in general in \cite{ACC} and this paper is an explicit illustration of the process on the AL model. 

Another outstanding question at this stage is how to deal with our time-dependent boundary conditions in the scheme of the so-called nonlinear mirror image. The latter has been well-known since the early papers \cite{H,BT1,BT2,T} where it appeared in connection with the use of B\"acklund transformation together with a folding procedure. However, up to now, this scheme has only been use in connection with the time-independent boundary zero curvature equation of the form \eqref{eq:ZCB1}. We develop the time-dependent nonlinear mirror image method in two steps. In the next subsection, we present the first step in implementing this approach with the view of constructing explicit solutions: we embed \eqref{time_dep_BC} into a B\"acklund transformation scheme coupled with a folding prescription. Then, in Section \ref{mirror}, this is used in conjunction with the standard ISM framework to construct explicit solutions of our AL model on the half-line with integrable boundary conditions.

\subsection{Auxiliary problem, B\"acklund transformation and nonlinear mirror image method}\label{Aux}

The zero curvature equations presented in Proposition \ref{pro:ZC} provide the consistency relations for the following auxiliary problem for the $2\times 2$ matrices $\Phi(n,t,z)$\footnote{In the auxiliary problem approach that we set up now, the Lax pair depends on $t$ through the fields in the usual fashion but we prefer to keep the Hamiltonian notation without mentioning $t$ explicitely in the Lax pair, for continuity of notations.}
\bea
&&\Phi(j+1,t,z)=\cL(j,z)\,\Phi(j,t,z)\,,\ \text{for } j=0,1,\dots  \label{eq:aux1}\\
&&\partial_t \Phi(j,t,z)=\cA(j,z)\,\Phi(j,t,z)\,,\ \text{for } j=0,1,\dots
\eea
with the constraint 
\bea
&&\Phi(0,t,z)=\rho(z) K^-(z)\,\Phi(0,t,\tau(z))\,.\label{eq:aux4}
\eea
Here, $\rho(z)$ is a function of the spectral parameter independent of $n$ and $t$,
chosen such that 
\be
\rho(z)\rho(\tau(z)) K^-(z) K^-(\tau(z))=\1\,.
\ee
 The existence of such function 
can be shown but in the following we do not need its explicit expression.
We now show how this auxiliary problem arises from a B\"acklund transformation by using the nonlinear mirror image method.
We start from two solutions $(Q_j,R_j)$ and $(\widetilde Q_j, \widetilde R_j)$ of the AL model on the full line ($j\in \ZZ$) such that
they are related by the following folding conditions
\be\label{eq:fold}
\widetilde Q_{j}=-\left(\frac{\beta}{\alpha}\right)^{1/2+j} Q_{-j-1}\,,~~\widetilde R_{j}=-\left(\frac{\alpha}{\beta}\right)^{1/2+j} R_{-j-1}\;.
\ee
It is straightforward to show that this transformation leaves the equations of motion of the AL model invariant.
We have also the following properties on the Lax pair due to \eqref{eq:fold}
\bea
&&\label{symmetry_cL}
\widetilde{\cL}(j,z)=J^{j+1}\, \cL^{-1}(-j-1,\tau(z))\,J^{-j}\;,\\
&&\label{symmetry_cA}
\widetilde{\cA}(j,z)=J^{j}\,\cA(-j,\tau(z))\,J^{-j}\;,
\eea
where
\be
 J=\begin{pmatrix}
\sqrt{\frac{\beta}{\alpha}} & 0\\
0 & \sqrt{\frac{\alpha}{\beta}}
\end{pmatrix}\;.
\ee
In particular, one gets 
\begin{equation}
 \widetilde \cA(0,z)=\cA(0,\tau(z))\;.\label{eq:syA1}
\end{equation}
Due to this symmetry on the Lax pair, we deduce that $\widetilde{\Phi}(j,t,z)$ and $J^{j}\, \Phi(-j,t,\tau(z))$ are eigenvectors of the same auxiliary problem. Therefore, there exists 
a matrix $M(z)$ independent of the time and of the position such that
\be
\widetilde{\Phi}(j,t,z)= J^{j}\, \Phi(-j,t,\tau(z))\ M(z)\;.\label{eq:symPhi}
\ee
We choose the normalisation of $\widetilde{\Phi}(j,t,z)$ such that $M(z)=\rho(z) \,\II$.
If we denote by $B(j,t,z)$ the matrix realising the B\"acklund transformation between both solutions, 
\begin{equation}
B(j,t,z)\widetilde \Phi(j,t,z)=\Phi(j,t,z)\;, \ \ \text{for }j\in\ZZ \label{eq:bFF}
\end{equation}
then it must satisfy relations \eqref{eq:BAper1}-\eqref{eq:BAper2} for all $j\in \ZZ$. 
As in the periodic case equation \eqref{eq:BAper2} taken at one particular value $j_0$ implies \eqref{eq:BAper2} for all $j\in\ZZ$.
For $j_0=0$, one gets explicitly 
\begin{equation}
\partial_t B(0,t,z) = \cA(0,z) B(0,t,z)- B(0,t,z) \widetilde \cA(0,z)\;.\label{eq:B1T}
\end{equation}
By using \eqref{eq:syA1}, we see that relation \eqref{eq:B1T} is similar to \eqref{eq:ZCB1K} and we can choose consistently
\begin{equation}
 B(0,t,z) = K^-(z)\;. \label{eq:BK}
\end{equation}
This condition completely determines the B\"acklund transformation (see below for the case studied here). 
Then, with the choice \eqref{eq:BK} of boundary condition for the B\"acklund matrix, relation \eqref{eq:bFF} for $j=0$
becomes (with the symmetry relation \eqref{eq:symPhi})
\begin{equation}
\rho(z) K(z) \Phi(0,t,\tau(z))=\Phi(0,t,z)\;.
\end{equation}
This is nothing but relation \eqref{eq:aux4} and we conclude that the solution $Q_j$ and $R_j$ satisfy the desired boundary conditions encoded in \eqref{eq:ZCB1K}. This shows that the folding procedure coupled with the B\"acklund transformation approach yields solutions of the equations of motion with the desired integrable boundary conditions. It turns out that it also implies the following symmetry relation on $B(j,t, z)$
\begin{lem}\label{lem:symB}
The B\"acklund matrix $B(j,t,z)$ in the folding procedure explained above satisfies
 \be
\label{symmetry_Bn}
B^{-1}(-j,t,\tau(z))=\rho(z)\rho(\tau(z)) J^{-j}\,B(j,t,z)\,J^{j}\;.
\ee
\end{lem}
\proof
By definition of $B(j,t,z)$, we have
\be
B(j,t, z)=\Phi(j,t, z)\, \widetilde{\Phi}^{-1}(j,t, z)\,.
\ee
From \eqref{eq:symPhi}, it implies that 
\be
\label{form_Bn}
B(j,t, z)=\frac{1}{\rho(z)}\Phi(j,t, z)\ \Phi^{-1}(-j,t,\tau( z))\,J^{-j}\,,
\ee
and \eqref{symmetry_Bn} is a consequence of \eqref{form_Bn}.
\endproof

The results of Lemma \ref{lem:bac} are still valid in this case, simply replacing $q_j,r_j$ by  $Q_j,R_j$.
By using condition \eqref{eq:BK}, one fixes the arbitrary parameters as
\begin{eqnarray}
&& y^{1}_0=x^{2}_0=0 \quad , \qquad g^{1}_0=\frac{b}{\alpha}\quad , \qquad  f^{2}_0=\frac{b}{\sqrt{\alpha\beta}}\;,\\
&& f^{1}_0= \sqrt{\frac{\alpha}{\beta}}g^{2}_0=-\sqrt{\frac{\alpha}{\beta}}\ \frac{cd}{y^{2}_0}=-\frac{\beta}{\alpha}\ \frac{cd}{x^{1}_0}
=\frac{ a\mp \sqrt{4cd(1-Q_0R_0)+a^2}) }{2}\;.
\end{eqnarray}
As mentioned previously, we have that $\det(\cL(j,z))=1$ which implies that $\det(B(j,t,z))$ is independent of $j$ (it is easy to see that on relation \eqref{eq:BAper1}).
Therefore, we get in particular 
\be\label{eq:Bin}
\det(B(+\infty,t,z))=\det(B(0,t,z))=\det(K^-(z))\;.
\ee
Note that these determinants are also time independent due to the tracelessness of $\cA$ and $\widetilde{\cA}$. To determine the form of 
$B(+\infty,t,z)$, we assume that the fields vanish at infinity (this could be shown following the same argument as in \cite{BH}).
Looking at the coefficients $z^2$ and $z^{-2}$ in relation \eqref{eq:Bin}, we deduce that
\be
f^2_\infty=\frac{f^1_\infty ab}{\sqrt{\alpha\beta}((f^1_\infty)^2-c d)} \quad\text{and}\quad g^1_\infty=\frac{f^1_\infty ab}{\alpha((f^1_\infty)^2-c d)}\;,
\ee
where 
\be
f^r_\infty=\lim_{n\to\infty}f^r_n\,,~~g^r_\infty=\lim_{n\to\infty}g^r_n\,,~~r=1,2\,.
\ee
In this case, we obtain that $B(+\infty,t, z)$ is in fact independent of $t$ and we have
\begin{equation}\label{eq:Binf}
 B(+\infty, z) =\begin{pmatrix}
                 \varphi(z)  & 0\\
                  0 & \varphi(\tau(z))
                \end{pmatrix}\quad\text{where}\qquad \varphi(z)=f^{1}_\infty  z +\frac{ab f^{1}_\infty }{\alpha((f^{1}_\infty)^2-cd) z}-\frac{cd\beta}{\alpha f^{1}_\infty z^3}\;.
\end{equation}
The coefficients of $z^0$ in relation \eqref{eq:Bin} gives a constraint for $f^{1}_\infty$ which reads
\begin{equation}\label{eq:f1inf}
 \big((f^{1}_\infty)^2+a f^{1}_\infty -cd\big) \big((f^{1}_\infty)^2-a f^{1}_\infty -cd\big)
 \left(  (f^{1}_\infty)^2-\frac{b}{\sqrt{\alpha\beta}} f^{1}_\infty  -c d  \right)
 \left(  (f^{1}_\infty)^2+\frac{b}{\sqrt{\alpha\beta}} f^{1}_\infty  -c d  \right)=0\;.
\end{equation}
For any solution $f^{1}_\infty$ of \eqref{eq:f1inf}, one gets that $B(+\infty,\tau(z))B(+\infty,z)=\frac{1}{\rho(z)\rho(\tau(z))}\II$. Then, in particular, one gets
\be\label{eq:rhophi}
\varphi(z)\varphi(\tau(z))=\frac{1}{\rho(z)\rho(\tau(z))}\;.\ee  
By using the result of the Lemma \ref{lem:symB}, one gets that $B(+\infty,\tau(z))B(-\infty,z)=\frac{1}{\rho(z)\rho(\tau(z))}\II$. Finally, we conclude that
\begin{equation}
 B(-\infty,z)=B(+\infty,z)\;.\label{eq:symB}
\end{equation}

We want to stress that in this paper, the choice of the value of $B(0,t,z)$ is dictated by the solution $K^-$ of the reflection equation. This is in contrast with previous approaches, e.g. \cite{BH}, where educated guesses for $B(0,t,z)$ are taken in order to produce the desired boundary conditions under the folding method. Our approach has an advantage in practice (no guess work) but it also has a deeper significance which is the underlying message in this paper: the Hamiltonian approach and the zero curvature approach to integrable boundary conditions are not separated topics but two faces of the same coin. This interplay is well-known for problems on the line but somehow was not as clear in the case with boundaries, despite the seminal work of Sklyanin \cite{Sk}. In particular, the reflection equation provides the reflection matrices that can be used as the boundary condition for the B\"acklund matrix that one uses to perform the nonlinear mirror image method.

\section{Application of the nonlinear mirror image method to the construction of solutions}\label{mirror}

In this section, we draw on the results of the previous section about the folding procedure associated to B\"acklund transformations in order to derive explicit solutions of the equations of motions with our integrable boundary conditions. To do so, we first need to review the ISM for the AL model (in our normalisation) on the full line. Then, the construction of solutions on the half-line is implemented as a special $\ZZ_2$ reduction realised by the B\"acklund matrix $B(j,t,z)$ derived above. For the sake of clarity, we focus on case $\alpha=\beta=\frac{1}{2}$ and $\gamma=-1$ in all this section. Then, one gets from now on $J=\II$, $\tau(z)=1/z$ and $\omega(z)=\frac{1}{2}(z-1/z)^2$.

After presenting the main results we need, we will further restrict our attention to the discrete NLS case $R_j=\nu Q_j^*$, $\nu=\pm 1$. As we are interested in soliton solutions, we will set $\nu=-1$.

\subsection{Review of the ISM for AL on the full lattice in the normalisation \eqref{normalisation_ell}}

In \cite{XF}, the ISM for the AL system in the normalisation \eqref{normalisation_ell} was presented, with an emphasis on the Riemann-Hilbert formulation of the inverse part. 
For our purposes, \ie the construction of explicit multisoliton solutions, the detailed results of \cite{APT} in the original normalisation will be more useful. To be able to use them, we first need to make the 
connection between the two normalisations as far as ISM is concerned. The best way is to realise that they area related by a gauge transformation as follows.

Let us work with the fields $Q_j$, $R_j$, $j\in\ZZ$.
We consider the auxiliary problem on the full lattice ($j\in \ZZ$) in the new normalisation,
\bea
&&\Phi^{new}(j+1,t,z)=\cL(j,z)\,\Phi^{new}(j,t,z)\,,\\
&&\partial_t \Phi^{new}(j,t,z)=\cA(j,z)\,\Phi^{new}(j,t,z)\,,
\eea
where
\begin{eqnarray}
&&{\cL}(j,z)=\frac{1}{N_j}\left(Z+W_j \right)\;,\\
&&\cA(j,z)=
i\omega(z)\sigma_3+i\sigma_3\Big(ZW_j - W_{j-1} Z - \frac{1}{2}W_j\ W_{j-1}- \frac{1}{2}W_{j-1}\ W_j   \Big)\;,
\end{eqnarray}
and,
\begin{equation}
N_j=\sqrt{1-Q_j R_j }\,,~~ Z=\begin{pmatrix}
z &0\\
0 &\frac{1}{z}
\end{pmatrix}
\,,~~ \ W_j=\begin{pmatrix}
              0&Q_j\\
              R_j & 0
             \end{pmatrix}\;.
\end{equation}
We also consider the auxiliary problem on the full line in the old normalisation, for $j\in \ZZ$,
\bea
&&\Phi^{old}(j+1,t,z)=U(j,z)\,\Phi^{old}(j,t,z)\,,\\
&&\partial_t \Phi^{old}(j,t,z)=V(j,z)\,\Phi^{old}(j,t,z)\,,
\eea
where
\begin{eqnarray}
&&U(j,z)=\left(Z+W_j \right)\;,\\
&&V(j,z)=
i\omega(z)\sigma_3+i\sigma_3\Big(ZW_j - W_{j-1} Z - W_j\ W_{j-1} \Big)\;,
\end{eqnarray}
One can check that the two are related by a gauge transformation of the following form (up to a possible normalisation constant, see below)
\be
\Phi^{old}(j,t,z)=F(j,t)\,\Phi^{new}(j,t,z)\,,
\ee
where $F$ is a scalar function satisfying
\be
F(j+1,t)=N_j\,F(j,t)\,,~~\partial_t F(j,t)=F(j,t)(V(j,z)-\cA(j,z))\,.
\ee
We fix $F$ by normalising it to $1$ as $j\to-\infty$ to get
\be
F(j,t)=\prod_{k=-\infty}^{j-1}N_k\,.
\ee
We will use this gauge transformation below to transfer the results of \cite{APT} to our setup. 

Let us now review ISM for AL. The initial sequences $Q_j|_{t=0}$, $R_j|_{t=0}$ are assumed to have finite $1$-norm where 
\be
||u||_1=\sum_{j\in\ZZ}|u_j|\,.
\ee
This ensures the desired analyticity properties below \cite{APT}. Also, they are assumed to be such that $N_j$ is defined and nonzero for all $j\in\ZZ$.
Define
\be
\Psi^{old/new}(j,t,z)=Z^{-j}e^{-i\omega(z)t\sigma_3}\,\Phi^{old/new}(j,t,z)
\ee 
and consider the two fundamental solutions $\Psi^{old/new}_\pm(j,t,z)$ normalised to $\1$ as $j\to\pm\infty$:
\be
\lim_{j\to\pm\infty}\Psi^{old/new}_\pm(j,t,z)=\1\,,
\ee
with corresponding Jost solutions $\Phi^{old/new}_\pm(j,t,z)$. The scattering matrix $S^{old/new}(z)$ is defined by
\be
\Phi^{old/new}_-(j,t,z)=\Phi^{old/new}_+(j,t,z)\,S^{old/new}(z)\,,~~|z|=1\,.
\ee
The gauge transformation between the two Lax pairs implies that
\be
\label{gauge}
\Phi^{old}_-(j,t,z)=F(j,t)\,\Phi^{new}_-(j,t,z)\,,~~\Phi^{old}_+(j,t,z)=\frac{F(j,t)}{F_\infty}\,\Phi^{new}_+(j,t,z)\,,
\ee
where
\be
F_\infty=\lim_{j\to\infty}F(j,t)\,.
\ee
It can be shown that $F_\infty$ is time-independent and hence is just a constant number. As a consequence, we find
\be
\label{relationS}
S^{old}(z)=F_\infty\, S^{new}(z)\,.
\ee
This is the key relation we need to relate the results of \cite{APT} with our setup.
One convenient consequence of the new normalisation of $\cL(j,z)$ is that 
\be
\det \Phi^{new}_\pm(j,t,z)=1\,,~~\det S^{new}(z)=1\,,~~|z|=1\,.\label{eq:det1}
\ee
This is of course consistent with the known fact \cite{APT} that $\det S^{old}(z)=F_\infty^2$ for $|z|=1$.
The analyticity properties in $z$ of the column vectors of $\Phi^{old/new}_\pm(j,t,z)$ and that of the entries of $S^{old/new}(z)$ are crucial for the implementation of the ISM. We see that the gauge transformation 
\eqref{gauge} and relation \eqref{relationS} have no consequence on these properties (domain of analyticity, location of zeros). The only consequences are to produce an overall factor $F_\infty$ between the so-called norming constants in the old and new normalisation and to change the normalisation of the scattering coefficients which now reads (see \cite{XF})
\bea
\label{asymptotics1}
&&s_{11}(z)=\frac{1}{F_\infty}+O\left(z^{-2}\right)\,,~~z\to\infty\,,\\
\label{asymptotics2}
&&s_{22}(z)=\frac{1}{F_\infty}+O\left(z^{2}\right)\,,~~z\to 0\,.
\eea
 The change of normalisation of the norming constants can always be absorbed by redefining them. Therefore, in the following, we simply review the results of ISM that we need from \cite{APT} and assume that the overall constant $F_\infty$ has been absorbed in the norming constants. In particular, we now focus on the new normalisation and drop the superscript $new$ for conciseness.

Let us split the Jost solutions into column vectors and write
\be
\Phi_\pm(j,t,z)=\left(\Phi^L_\pm(j,t,z),\Phi^R_\pm(j,t,z) \right)\,,~~S(z)=\begin{pmatrix}
s_{11}(z) & s_{12}(z)\\
s_{21}(z) & s_{22}(z)
\end{pmatrix}
\ee
One shows that $\Phi_+^L(j,t,z)$, $\Phi_-^R(j,t,z)$ and $s_{22}(z)$ are analytic functions of $z$ for $|z|< 1$ and continuous for $|z|\le 1$ while 
$\Phi_-^L(j,t,z)$, $\Phi_+^R(j,t,z)$ and $s_{11}(z)$ are analytic functions of $z$ for $|z|>1$ and continuous for $|z|\ge 1$.
We consider a finite number of simple zeros $z_n$, $n=1,\dots,P$ of $s_{22}(z)$ in the region $|z|> 1$ and a finite number of zeros $\overline{z}_n$, 
$n=1,\dots,\overline{P}$ of $s_{11}(z)$ in the region $|z|<1$.
At those zeros, the column vectors of $\Phi_\pm(j,t,z)$ are related by the so-called norming constants as follows
\bea
\label{def_bj}
&&\Phi_-^L(j,t,z_n)=b_n\,\Phi_+^R(j,t,z_n)\,,\\
\label{def_tbj}
&&\Phi_-^R(j,t,\overline{z}_n)=\overline{b}_n\,\Phi_+^L(j,t,\overline{z}_n)\,.
\eea
It will be convenient to introduce another set of norming constants defined by
\be\label{eq:defC}
C_n=\frac{b_n}{s^{'}_{11}(z_n)}\,,~~\overline{C}_n=\frac{\overline{b}_n}{s^{'}_{22}(\overline{z}_n)}\,.
\ee

Equipped with the scattering data, one can implement the inverse part of the method to obtain reconstruction formulas 
for the solution of the AL system on the full lattice. 
There is an inherent symmetry on the discrete data. It comes from the fact that the Lax matrices $\cL(j,z)$ and $\cA(j,z)$ satisfy the following symmetry relation
\be
\cM(j,z)=-\sigma_3\,\cM(j,-z)\,\sigma_3\,,~~\cM=\cL,\cA\,.
\ee
This implies
\be
\Phi_\pm(j,t,z)=(-1)^j\sigma_3\,\Phi_\pm(j,t,-z)\,\sigma_3
\ee
and
\be
S(z)=\sigma_3\,S(-z)\,\sigma_3\,.
\ee
As a consequence, the zeros of $s_{11}(z)$ (resp. $s_{22}(z)$) come in pairs $\pm z_n$ and hence $P=2\kappa$ for some integer $\kappa\ge 0$ (resp. $\pm \overline{z}_n$, $\overline{P}=2\overline{\kappa}$). 
One can show that the norming constant $C_n$ (resp. $\overline{C}_n$) associated to $z_n$ (resp. $\overline z_n$) is equal to the norming constant associated to $-z_n$ (resp. $-\overline z_n$). 
Equations (3.2.102)-(3.2.104) of \cite{APT} exploit this symmetry and can be used to derive a nice compact formula for the pure soliton case associated to (half of) the discrete data 
$\{z_1,\dots,z_\kappa;C_1,\dots,C_\kappa;\overline{z}_1,\dots,\overline{z}_{\overline{\kappa}};\overline{C}_1,\dots,\overline{C}_{\overline{\kappa}} \}$. 
After some calculations, we find\footnote{We shifted $j-1\to j$ compared to \cite{APT} when expressing $Q_j$ as we find it more convenient.}
\bea
\label{soliton1}
&&Q_{j}(t)=-2(1\dots 1)\; \overline{\mu}_j^{-1}(t)\; \begin{pmatrix}
\overline{C}_1\ \overline{z}_1^{2j}\  e^{-2i\omega(\overline{z}_1)t}\\
\vdots\\
\overline{C}_{\overline{\kappa}}\ \overline{z}_{\overline{\kappa}}^{2j}\  e^{-2i\omega(\overline{z}_{\overline{\kappa}})t}
\end{pmatrix}\,,\\
&&R_{j}(t)=2(1\dots 1)\; \mu_j^{-1}(t)\; \begin{pmatrix}
C_1\  z_1^{-2j-2}\ e^{2i\omega(z_1)t}\\
\vdots\\
C_\kappa\  z_\kappa^{-2j-2}\ e^{2i\omega(z_\kappa)t}
\end{pmatrix}\,,
\eea
where the $\kappa\times\kappa$ matrix $ \mu_j(t)$ has the following entries
\be
( \mu_j(t))_{nl}=\delta_{nl}-4\sum_{k=1}^{\overline{\kappa}}\frac{C_n\overline{C}_k\ z_n^{-2j}\; \overline{z}_k^{2(j+1)}\ e^{2i(\omega(z_n)-\omega(\overline{z}_k))t}}{(z_n^2-\overline{z}_k^2)
(\overline{z}_k^2-z_l^2)}\,,~~n,l=1,\dots,\kappa\,,
\ee
and the $\overline{\kappa}\times\overline{\kappa}$ matrix $\overline{\mu}_j(t)$ has the following entries
\be
\label{soliton4}
(\overline{\mu}_j(t))_{nl}=\delta_{nl}-4\sum_{k=1}^{\kappa}\frac{C_k\overline{C}_n\ \overline{z}_n^{2(j+1)}\; z_k^{-2j}\ e^{2i(\omega(z_k)-\omega(\overline{z}_n))t}}
{(\overline{z}_n^2-z_k^2)(z_k^2-\overline{z}_l^2)}\,,~~n,l=1,\dots,\overline{\kappa}\,.\ee

From now on, we set $R_j=- Q_j^*$. This leads to an additional symmetry on the Lax pair which in turns implies a symmetry on the scattering data which should be implemented in the above formulas. In short we have, 
\be
\label{DNLS_reduction}
\cM(j,z)=\sigma\,\cM^*(j,\frac{1}{z^*})\,\sigma^{-1}\,,~~\cM=\cL,\cA\,,~~
\sigma=\begin{pmatrix}
	0 & 1\\
	-1 & 0	
\end{pmatrix}\,.
\ee
It implies
\be
\Phi_\pm(j,t,z)=\sigma^{-1}\,\Phi^*_\pm(j,t,\frac{1}{z^*})\,  \sigma\,,
\ee
\be
S(z)=\sigma\,S^*(\frac{1}{z^*})\,  \sigma^{-1}\,,
\ee
and, on the discrete data, $\kappa=\overline{\kappa}$ and
\be
\overline z_n=\frac{1}{z_n^*}\quad\text{and}\quad \overline{C}_n=\frac{C_n^*}{(z_n^*)^2}\,.
\ee
With these, we can restrict our attention to \eqref{soliton1} and \eqref{soliton4}.

Under this reduction, we can use the trace formulas (3.2.89) of \cite{APT} in the pure soliton case to obtain the following useful explicit form of $s_{11}(z)$, $s_{22}(z)$, suitably normalised to our setup (see in particular \eqref{asymptotics1}-\eqref{asymptotics2})
\be
\label{formula_s}
s_{11}(z)=\frac{1}{F_\infty}\prod_{j=1}^{\kappa}\frac{z^2-z_j^2}{z^2-(z_j^*)^{-2}}\,,~~
s_{22}(z)=\frac{1}{F_\infty}\prod_{j=1}^{\kappa}|z_j|^4\frac{z^2-(z_j^*)^{-2}}{z^2-z_j^2}\,.
\ee
Since $\det S(z)=1$ when $|z|=1$ and $F_\infty>0$ in the DNLS reduction with $\nu=-1$, this fixes the value of $F_\infty$ completely in terms of the scattering data as\footnote{This does not seem to agree with the formula given in \cite{BHw} for the constant $C_{-\infty}$ which should correspond to our $F_\infty^2$.}
\be
F_\infty=\prod_{j=1}^{\kappa}|z_j|^2\,.
\ee

\subsection{Symmetry from the mirror image procedure}\label{sec:cons}

The important idea behind the mirror image method is that one can obtain solutions of an integrable PDE on the half-line with certain (integrable) boundary conditions by considering solutions of the full line problem associated 
to scattering data with a special symmetry. One way to obtain the desired symmetry on the scattering data is by interpreting the boundary conditions as arising from a well chosen B\"acklund transformation relating a solution 
to its ``mirror image''. This is what we have done above by constructing $B(j,t,z)$ and we use it now to derive the symmetry of the scattering data. 

We first consider the continuous scattering data
\begin{prop}\label{pro:Ss}
The following relation holds on the Jost solutions
\be\label{eq:jos}
\Phi_+(j,t,z)B(+\infty,z)  =B(j,t,z)\,\Phi_-(-j,t,\frac{1}{z})\,.
\ee
As a consequence, the scattering matrix satisfies
\be
\label{folding_sym_S}
S^{-1}(z)=B(+\infty,z)\,S(\frac{1}{z})\ B^{-1}(+\infty,z)\;.
\ee
\end{prop}
\proof
Let us denote by $\widetilde \Phi_\pm(j,t,z)$ the Jost solutions of the auxiliary problem with the mirror image solutions.
$\Phi_\pm(j,t,z)$ and $B(j,t,z) \widetilde \Phi_\pm(j,t,z)$ are solutions of the same auxiliary problem. Then, there exist
matrices $N_{\pm}(z)$ independent of time and position such that
\begin{equation}
 \Phi_\pm(j,t,z)=B(j,t,z) \widetilde \Phi_\pm(j,t,z)\ N_{\pm}(z)\;.
\end{equation}
By taking the limit $j\rightarrow \pm\infty$ in the previous relation and by using the asymptotic of the Jost solutions, one gets
\be
B(\pm \infty,z) \ N_{\pm}(z)=\II\;.
\ee
Then
\be\label{eq:phk}
\Phi_\pm(j,t,z)B(\pm \infty,z) =B(j,t,z) \widetilde \Phi_\pm(j,t,z)\;.
\ee
Because of the symmetry of the Lax pair \eqref{symmetry_cL}-\eqref{symmetry_cA}, $\widetilde \Phi_+(j,t,z)$ and $\Phi_-(-j,t,1/z)$ are also eigenvectors of the 
same auxiliary problem (we recall $J=\II$) and one gets
\begin{equation}\label{eq:ssd}
 \widetilde \Phi_+(j,t,z)= \Phi_-(-j,t,1/z) \;.
\end{equation}
It is easy to see that the normalisation in the previous relation is the identity by taking the limit $j\rightarrow \infty$.
Relations \eqref{eq:phk} and \eqref{eq:ssd} lead to relation \eqref{eq:jos} of the proposition.
Since $\Phi_-(j,t,z)=\Phi_+(j,t,z)\,S(z)$ and by using \eqref{eq:jos} by replacing $j\rightarrow -j$ and $z\rightarrow 1/z$, we obtain 
\be
\II=B(-\infty,z) S(\frac{1}{z}) B^{-1}(+\infty,z) S(z)\;.
\ee
From this and \eqref{eq:symB}, one gets \eqref{folding_sym_S} as desired.
\endproof
As an important consequence of the Proposition \ref{pro:Ss}, one gets
\be
\label{sym_a}
s_{11}(1/z)=s_{22}(z)\,,
\ee
and
\be
s_{12}(z)=f(z)s_{12}(1/z)\,,~~s_{21}(z)=f(1/z)s_{12}(1/z)\,,
\ee
where
\be
\label{def_f}
f(z)=-\frac{\varphi(z)}{\varphi(1/z)}\,.
\ee
We turn to the discrete data. The symmetry relation \eqref{sym_a} implies that $\kappa=\overline{\kappa}$ and without loss of generality
\be
\label{folding_sym_zj}
\overline{z}_n=\frac{1}{z_n}\,,~~n=1,\dots,\kappa\,.
\ee
Note that the reduction to $\kappa=\overline{\kappa}$ would hold even if we did not consider the DNLS reduction of the AL system. It is the result of the folding symmetry only. The same holds true for the folding symmetry on the scattering data that we discuss below. That being said, since the focus of this section is on DNLS, let us examine the effect of the DNLS reduction on the boundary data before we proceed to showing the effect of the folding symmetry on the discrete data.

\begin{lem}
	\label{lemmaDNLS}
Under the DNLS reduction with $\nu=-1$, the boundary parameters $a,b,c,d$	satisfy
\be
a,b\in\RR\,,~~c= d^*\,,
\ee
and the function $\varphi(z)$ satisfies
\be
\varphi(z)=\varphi^*(z^*)\,.
\ee
In particular, $f^1_\infty\in\RR$.
\end{lem}
\prf
The reduction also applies to $k^-(z)$ and $B(+\infty,z)$ which must therefore satisfy the symmetry condition \eqref{DNLS_reduction}. Direct calculation yields the stated results.
\finprf

\begin{prop}
Under the folding reduction, the discrete data satisfies
\be
\overline{z}_n=\frac{1}{z_n}\,,~~
C_n\,\overline{C}_n=-\ \frac{\varphi(1/z_n)}{(z_n\;s'_{11}(z_n))^2 \varphi(z_n)}\,,~~n=1,\dots,\kappa\,,
\ee
where $\varphi(z)$ is given by (see \eqref{eq:Binf}, taking into account the DNLS reduction)
\be
\label{poly}
\varphi(z)=f^{1}_\infty  z +\frac{2ab f^{1}_\infty }{((f^{1}_\infty)^2- |d|^2) z}-\frac{|d|^2}{ f^{1}_\infty z^3}\,,
\ee
and $f^1_\infty$ is a root of the following polynomial (see \eqref{eq:f1inf})
\begin{equation}
\big(((f^{1}_\infty)^2-|d|^2)^2-a^2 (f^{1}_\infty)^2\big) 
\left( ( (f^{1}_\infty)^2-|d|^2)^2-4b^2 (f^{1}_\infty)^2 \right)=0\;.
\end{equation}
\end{prop}
\proof
Recalling \eqref{def_bj}-\eqref{def_tbj} and using \eqref{folding_sym_zj} we have
\bea
&&\Phi_-^L(j,t, z_n)=b_n\,\Phi_+^R(j,t, z_n)\,,\\
&&\Phi_-^R(j,t,1/z_n)=\overline{b}_n\,\Phi_+^L(j,t,1/z_n)\,.
\eea
From relations \eqref{eq:Binf} and \eqref{eq:jos}, we obtain
\bea
&&\Phi_+^L(j,t,z)\varphi(z)=B(j,t,z)\,\Phi^L_-(-j,t,1/z)\,,\\
&&\Phi_+^R(j,t,z)\varphi(1/z)=B(j,t,z)\,\Phi^R_-(-j,t,1/z)\,.
\eea
Using \eqref{symmetry_Bn} and \eqref{eq:rhophi}, we derive
\be
\Phi_-^R(j,t,1/z_n)= \frac{b_n \overline{b}_n\varphi(z_n)}{\varphi(1/z_n)}\,\Phi_-^R(j,t,1/z_n)
\ee
which, together relations \eqref{eq:defC}, \eqref{sym_a}, gives the relation on the norming constants. The rest of the proposition follows from Lemma \ref{lemmaDNLS}.
\endproof

\subsection{Reflected solitons for the DNLS}

The three different symmetries on the scattering data (the one inherent to AL, the one coming from the DNLS reduction and the one associated to the folding reduction) are compatible, as we have seen. When they are imposed simultaneously, one can construct solutions of DNLS on the half-lattice with integrable boundary conditions using the solution on the full lattice and restricting to positive integers. 
More precisely, when all three symmetries hold, the numbers of zeros $P$ of $s_{22}$ is $P=2\kappa$ where $\kappa$ is itself an even integer, $\kappa=2k$, as a consequence of the folding symmetry. In that case, we also know that $\overline{P}=P=4k$. Therefore, the discrete data come in octets that produce one soliton each. Each octet is completely determined by one zero $z_n\in\CC$, $|z_n|>1$, and one norming constant $C_n\in\CC$ which are paired as follows, for $n=1,\dots,k$,
\bea
&&(z_n,C_n)\,,\\
&&(-z_n, C_n)\,,\\
&&(z_n^*,- \frac{\varphi^*(1/z_n)}{C_n^* (s'_{11}(z_n))^{*2} \varphi^*(z_n)})\,,\\
&&(-z_n^*,- \frac{\varphi^*(1/z_n)}{C_n^* (s'_{11}(-z_n))^{*2} \varphi^*(z_n)})\,,\\
&&(1/z_n,-\ \frac{\varphi(1/z_n)}{C_n (z_n\;s'_{11}(z_n))^2 \varphi(z_n)})\,,\\
&&(-1/z_n,-\ \frac{\varphi(1/z_n)}{C_n (z_n\;s'_{11}(-z_n))^2 \varphi(z_n)})\,,\\
&&(1/z_n^*,\frac{C_n^*}{(z_n^*)^2})\,,\\
&&(-1/z_n^*,\frac{C_n^*}{(z_n^*)^2})
\eea
The first four zeros in each octet correspond to $s_{11}(z)$ while the last four correspond to $s_{22}(z)$. The explicit formulas  
\eqref{formula_s} now take the form
\bea
&&s_{11}(z)=\frac{1}{F_\infty}\prod_{n=1}^{k}\frac{z^2-z_n^2}{z^2-(z_n^*)^{-2}}\frac{z^2-(z_n^*)^2}{z^2-(z_n)^{-2}}\,,\label{eq:s11b}\\
&&s_{22}(z)=\frac{1}{F_\infty}\prod_{n=1}^{k}|z_n|^8\frac{z^2-(z_n^*)^{-2}}{z^2-z_n^2}\frac{z^2-(z_n)^{-2}}{z^2-(z_n^*)^2}\,,
\eea
with
\be
F_\infty=\prod_{j=1}^{k}|z_j|^4\,.
\ee
The fact that each zero and its opposite have the same norming constant is intrinsic to the model as already discussed, and this has already been taken into account when deriving the explicit formulas \eqref{soliton1}-\eqref{soliton4}. Hence here, all we have to do is to take account the additional symmetries yielding $\kappa=2k$ into these formulas. 

\begin{prop}\label{form_solitons}
The $k$-soliton solution of DNLS on the half-infinite lattice with the new (time-dependent) boundary condition
\be
Q_{-1}=Q_1+(aQ_1+2bQ_0)\frac{a\pm\sqrt{a^2+4|d|^2(1+|Q_0|^2)}}{2|d|^2(1+|Q_0|^2)}\,,
\ee 
is determined by $k$ complex numbers $\zeta_1,\dots,\zeta_k$ with $|\zeta_j|>1$ and $k$ complex numbers $D_1,\dots,D_k$ and reads
\bea
&&Q_{j}(t)=-2(1\dots 1)\; \overline{\mu}_j^{-1}(t)\; \begin{pmatrix}
	\overline{C}_1\ \overline{z}_1^{2j}\  e^{-2i\omega(\overline{z}_1)t}\\
	\vdots\\
	\overline{C}_{2k}\ \overline{z}_{2k}^{2j}\  e^{-2i\omega(\overline{z}_{2k})t}
\end{pmatrix}\,,~~j\ge -1\,,
\eea
where the $2k\times 2k$ matrix $ \overline{\mu}_j(t)$ reads
\be
(\overline{\mu}_j(t))_{nl}=\delta_{nl}-4\sum_{p=1}^{2k}\frac{C_p\overline{C}_n\ \overline{z}_n^{2(j+1)}\; z_p^{-2j}\ e^{2i(\omega(z_p)-\omega(\overline{z}_n))t}}
{(\overline{z}_n^2-z_p^2)(z_p^2-\overline{z}_l^2)}\,,~~n,l=1,\dots,2k\,.
\ee
with the following conventions
\bea
&&z_n=\begin{cases}
	\zeta_n\,,~~n=1,\dots,k\,,\\
	\zeta_{n-k}^*\,,~~n=k+1,\dots,2k\,,
\end{cases}
\overline{z}_n=\begin{cases}
	1/\zeta_n\,,~~n=1,\dots,k\,,\\
	1/\zeta_{n-k}^*\,,~~n=k+1,\dots,2k\,,
\end{cases}\\
&&C_n=\begin{cases}
	D_n\,,~~n=1,\dots,k\,,\\
	- \frac{\varphi^*(1/\zeta_{n-k})}{D_{n-k}^* (s'_{11}(\zeta_{n-k}))^{*2} \varphi^*(\zeta_{n-k})}\,,~~n=k+1,\dots,2k\,,
\end{cases}
\overline{C}_n=\begin{cases}
	-\ \frac{\varphi(1/\zeta_n)}{D_n (\zeta_n\;s'_{11}(\zeta_n))^2 \varphi(\zeta_n)}\,,~~n=1,\dots,k\,,\\
	\frac{D_{n-k}^*}{(\zeta_{n-k}^*)^2}\,,~~n=k+1,\dots,2k\,.
\end{cases}
\eea
We recall that $\varphi$ is given by \eqref{poly} and $s_{11}$ by \eqref{eq:s11b}.
\end{prop}
The appearance of the discrete data in octets for the AL model with integrable boundary conditions was first shown in \cite{BHw} and then in \cite{BH} in the case of Robin boundary conditions. In our case, the entirety of the effect of our time-dependent boundary conditions is encoded in the function $\varphi(z)$ or, alternatively, in the function $f(z)$ in \eqref{def_f}. This is in line with the results of \cite{BH} for the Robin case. In fact, we can reproduce the known Robin case by choosing $c=d=0$ and for instance $a=-1$, $b=1/2\chi$: our function $f(z)$ in \eqref{def_f} becomes 
\be
f(z)=\frac{z\chi(f^1_\infty)^2-1/z}{z-\chi(f^1_\infty)^2/z}
\ee
which consistently reproduces the function $f(z)$ of \cite{BH} (eq.(7.12)) where our $(f^1_\infty)^2$ plays the role of $p_\infty$ in that paper. In our more general case, a thorough analysis of all the possible values that $f^1_\infty$ can take among the roots of \eqref{poly}, analogously to Corollary 6.4 of \cite{BH}, would be required to classify all the possible scenarios of solutions that one can construct using the mirror image method. We do not perform this analysis here as the number of cases to consider is much larger than in the Robin case. This technical point does not affect the significance of the results we have obtained. In particular, in the explicit examples to follow, we simply fix compatible numerical values of $f^1_\infty$ and $a,b,c,d$ to produce plots of solitons being reflected by our boundary conditions. 

In Figure 1, we present such a reflected solution in the one-soliton case. Left plots show a one-soliton being reflecting off the boundary at $x=-1$. We allowed $x$ to be real-valued but highlighted integer values in solid black curves. Right plots show contour plots of the one-soliton being reflected as well as the image soliton on the other side of the boundary (the black vertical line). For comparison, we display 3 types of boundary conditions: our time-dependent case and two particular cases of it which were previously known (Robin and Dirichlet).

\begin{figure}[htp]
\begin{center}
	\begin{minipage}[h]{6cm}
		\label{figa}
		\includegraphics[scale=0.4]{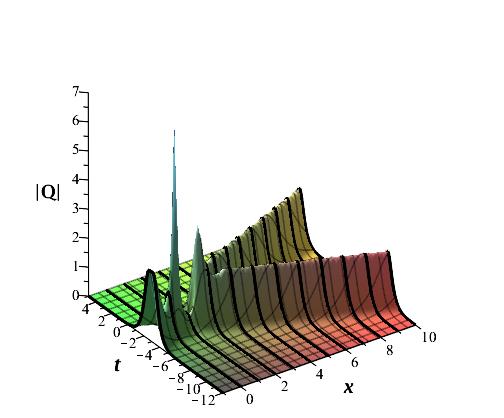}
$a)$ New boundary conditions: $a=1$, $b=-1.7$, $d=1.1$.
	\end{minipage}
	\qquad\qquad\qquad
	\begin{minipage}[h]{6cm}
		\includegraphics[scale=0.4]{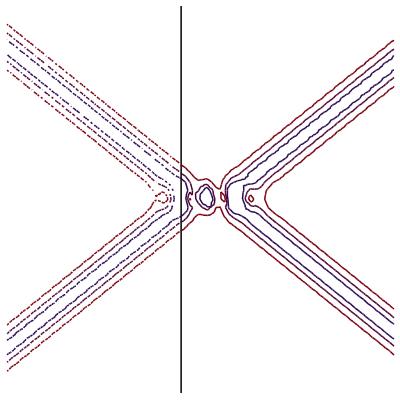}
$b)$ Contour plot with the mirror soliton.
	\end{minipage}

	\begin{minipage}[h]{6cm}
		\includegraphics[scale=0.4]{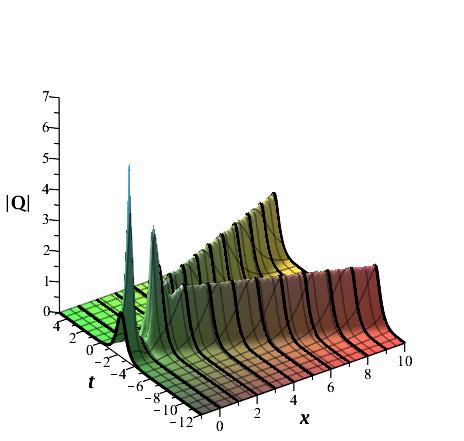}
$c)$ Robin boundary conditions: $a=1$, $b=-1.7$, $d=0$.
	\end{minipage}
	\qquad\qquad\qquad
\begin{minipage}[h]{6cm}
	\includegraphics[scale=0.4]{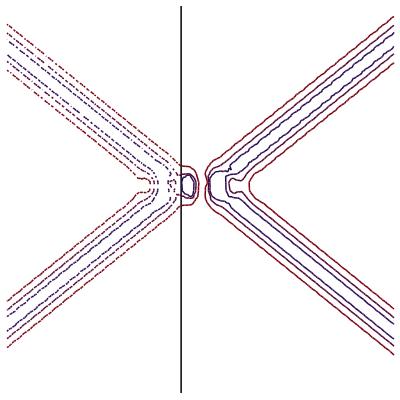}
$d)$ Contour plot with the mirror soliton.
\end{minipage}
	
	\begin{minipage}[h]{6cm}
	\includegraphics[scale=0.4]{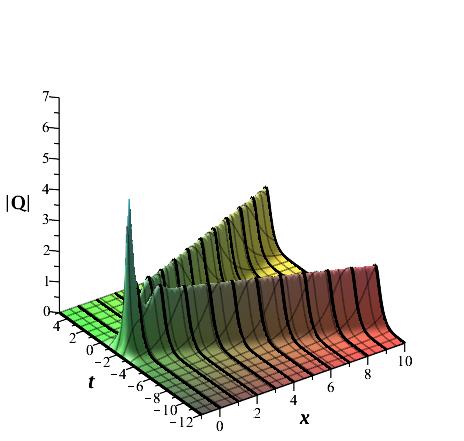}
$e)$ Dirichlet boundary conditions with $a=1$, $b=0$, $d=0$.
\end{minipage}
	\qquad\qquad\qquad
\begin{minipage}[h]{6cm}
	\includegraphics[scale=0.4]{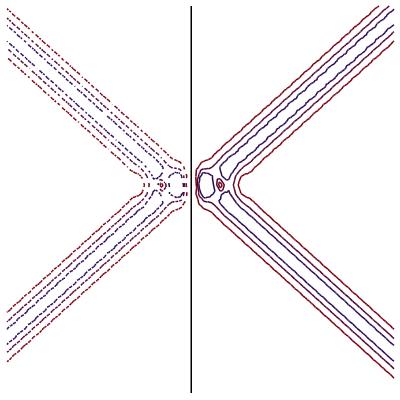}
$f)$ Contour plot with the mirror soliton.
	\label{figure1}
\end{minipage}
\caption{Time-dependent (top), Robin (middle) and Dirichlet (bottom) boundary conditions. Parameters of the soliton solution: $\zeta_1=0.6+1.9i$, $D_1=0.1$. }
\end{center}
\end{figure}

\section{Conclusions and outlook}

The main message illustrated in this paper is that the Hamiltonian approach and the zero curvature approach to integrable boundary conditions are not separated topics but two faces of the same coin. This interplay is well-known for problems on the line but somehow was not as clear in the case with boundaries, despite the seminal work of Sklyanin \cite{Sk}. The two aspects inform each other, as we have shown in detail with the AL model in this paper. For instance, the choice of the value of the B\"acklund matrix $B(j,t,z)$ at $j=0$ is dictated by the solution of the reflection equation that we consider. Taking non-diagonal solutions of the reflection equation, we showed that new, time-dependent, boundary conditions arise. For the first time, we then developed the nonlinear mirror image method for such boundary conditions and constructed explicit soliton solutions. 
A study of a continuous model, such as the nonlinear Schr\"odinger (NLS) equation, along the lines of the present work would be desirable to investigate what kind of more general integrable boundary conditions than the standard Robin boundary conditions can be imposed. This is currently under investigation. We note that the Hamiltonian aspects of this question, for the (vector) NLS equation, have already been discussed in \cite{DFR}. However, the full connection to zero curvature representation and to the nonlinear mirror method for the construction of solutions remains an open problem. There are alternative discretizations of the NLS model \cite{SE,KR} for which a study of boundary conditions related to (non diagonal) reflection matrices of the rational type along the lines of the present paper are an interesting open problem. Such an investigation would provide a complementary point of view on the NLS with time-dependent boundary conditions as continuous limit.

The results illustrated here rely on the general procedure described in \cite{ACC} which only involves fundamental features of integrable systems, such as $r$-matrix and Lax matrix. Therefore, there is no obstacle in principle to apply the same ideas for other models, with the important proviso that the model allows for a natural folding ($\ZZ_2$) symmetry in order to implement the nonlinear mirror image method. The problem of understanding the analog of the mirror image method for models that do not possess a natural $\ZZ_2$ symmetry is completely open and rather fascinating. So far, the only alternative to discuss such models (e.g. KdV) on the half-line with boundary conditions is to use the so-called unified transform \cite{Fok1}. It is a completely open problem to investigate integrable time-dependent boundary conditions of the kind we found in that setup. 

Finally, the quantization of our results is a natural question. The quantum Ablowitz--Ladik model with periodic boundary conditions \cite{Kul} has been well studied. Regarding boundary conditions, the algebraic Bethe ansatz was considered in \cite{CD} for diagonal reflection matrices. A more recent study of the $q$-boson model, related to the quantum Ablowitz--Ladik model, with boundary conditions can be found in \cite{DF} where the reflection matrices were also chosen to be diagonal (equal to the identity actually). The investigation of the quantization of our results is currently under way.

\paragraph{Acknowledgment:} N. Cramp\'e acknowledges the hospitality 
of the School of Mathematics, University of Leeds where this work was completed. N. Cramp\'e's visit was partially supported by the Research Visitor Centre of the School of Mathematics.

\end{document}